\documentstyle[12pt]{article}
\date{}
\begin{document}

\title{The Spectrum of the Nucleons and the
Strange Hyperons and Chiral
Dynamics}

\author{L. Ya. Glozman$^{1,2}$ and D.O. Riska$^{3,4}$}
\maketitle

\centerline{\it $^1$Institute for Theoretical Physics,
University of Graz, 8010 Graz, Austria}

\centerline{\it $^2$Alma-Ata Power Engineering Institute,
480013 Alma-Ata, Kazakhstan}

\centerline{\it $^3$Institute for Nuclear Theory, University
of Washington, Seattle, WA 98195}

\centerline{\it $^4$Department of Physics, University of Helsinki,
00014 Finland}

\setcounter{page} {0}
\vspace{1cm}

\centerline{\bf Abstract}
\vspace{0.5cm}

The spectra of the nucleons, $\Delta$ resonances and the
strange hyperons are well described by the constituent
quark model if in addition to the
harmonic confinement potential
the quarks are assumed to interact by exchange of the $SU(3)_F$
octet
of pseudoscalar mesons, which are the Goldstone
bosons associated with the hidden
approximate chiral symmetry of QCD.
In its $SU(3)_F$ invariant approximation
the pseudoscalar exchange interaction splits the multiplets
of $SU(6)_{FS}\times U(6)_{conf}$ in the spectrum to multiplets
of $SU(3)_F\times SU(2)_S\times U(6)_{conf}$.
The position of these multiplets differ in the baryon sectors
with different strangeness
because of  the mass splitting
of the pseudoscalar octet and the different constituent masses of the
u,d and s quarks that breaks $SU(3)_F$ flavor symmetry.
A description of the whole spectrum, to an accuracy of $\simeq$ 4\%
or better,
is achieved if one matrix element of the boson interaction for
each oscillator shell
is extracted from the empirical mass splittings.
The ordering of the positive and negative parity states
moreover agrees with the empirical one in all sectors of the
spectrum. A discussion of the conceptual basis of the model
and its various phenomenological ramifications is presented.\\

Preprint DOE/ER/40561-187-INT95-16-01
hep-ph 9505422
Submitted to Physics Reports 7.05.1995

\newpage
\normalsize

\centerline{\bf 1. Introduction}
\vspace{1cm}

The spectra of the confirmed states of the nucleon and the
$\Lambda$ hyperon separate into a low energy sector of well
separated states
without nearby parity partners, and a high energy sector
with an increasing number of near parity doublets. A natural
interpretation of this feature
is that the approximate chiral symmetry of QCD is realized in
the hidden Nambu-Goldstone mode at low excitation (and
temperature) and in the
explicit Wigner-Weyl mode at high excitation.\\

The hidden mode of chiral symmetry
is revealed by the existence of the octet of  pseudoscalar
mesons of low mass, which represent the associated approximate
Goldstone bosons. The $\eta'$ (the $SU(3)$-singlet) decouples from the original
nonet because of the $U(1)$ anomaly \cite{WE,THO}.
Another consequence of the spontaneous breaking of the approximate
chiral symmetry of QCD is that the valence quarks acquire
their dynamical or constituent mass
\cite{WEIN,MAG,SHU,DIP,DIA}
through their interactions with the collective excitations of
the QCD vacuum-
the quark-antiquark excitations and the instantons. The origin of this
dynamical generation of the constituent quark mass is closely related
to the origin of the pseudoscalar Goldstone excitations.
Thus according to the two-scale picture
of Manohar and Georgi \cite{MAG}
the appropriate effective degrees of freedom for
the 3-flavor QCD
at distances beyond that of spontaneous chiral symmetry breaking
(0.2--0.3 fm), but within that of the confinement scale
$\Lambda_{QCD}^{-1} \simeq 1 fm$,
should be the
constituent quarks with internal structure,
and the chiral meson fields.\\

In line with this we have recently suggested \cite{GLO1,GLO2}
that beyond the chiral symmetry
spontaneous breaking scale a baryon should be considered
as a system of three constituent quarks
with an effective quark-quark interaction that is formed
of a central confining part, assumed to be harmonic, and a chiral
interaction that is mediated by the octet of pseudoscalar mesons
between the constituent quarks.\\

Even in its simplest $SU(3)_F$ invariant form this boson
exchange interaction between the constituent quarks leads to a remarkably
good description of the whole hitherto measured spectrum of
the nucleon, $\Delta$ resonance and $\Lambda$ hyperon
\cite{GLO1,GLO2}. We here develop this model in more detail, with
full account of the $SU(3)_F$ breaking caused by the mass splitting
of the pseudoscalar octet and different constituent masses of the
u,d and s quarks, and show that it provides a very
satisfactory representation of the known parts of the
spectra of the $\Sigma$, $\Xi$ and
$\Omega$ hyperons as well.\\

	The simplest representation of the most important
component of the interaction of the constituent quarks that is mediated
by the octet of pseudoscalar bosons in the $SU(3)_F$
invariant limit is

$$H_\chi\sim -\sum_{i<j}V(\vec r_{ij})
\vec \lambda^F_i \cdot \vec \lambda^F_j\,
\vec
\sigma_i \cdot \vec \sigma_j.\eqno(1.1)$$
Here the $\{\vec \lambda^F_i\}$:s are flavor $SU(3)$ Gell-Mann
matrices and the
$i,j$ sums run over the constituent quarks. The interaction
potential $V(r)$ will have the usual Yukawa behavior at long
range, but at short range  behaves as a smeared version
of the $\delta$ function term in the Yukawa
interaction for pseudoscalar exchange.\\

If the only interaction between
the quarks were the flavor- and spin- independent
harmonic confining interaction
the baryon spectrum would be organized
in multiplets of the symmetry group $SU(6)_{FS}\times U(6)_{conf}$,
as the symmetry of the 3-quark states in the harmonic oscillator
basis is $U(6)_{conf}$
and the permutational $SU(6)_{FS}$ symmetry is uniquely determined
by the $U(6)_{conf}$ symmetry by the Pauli principle.
In this case the baryon masses  would
be determined solely by the orbital structure and by the constituent
quark masses and the spectrum would be
organized in an alternating sequence of positive and negative
parity states.
This multiplet structure of the
spectrum is broken by the
interaction (1.1) between the constituent quarks,
and in the first order perturbation in the $SU(3)_F$
symmetric approximation
for the interaction
the multiplet structure
is then that of the group $SU(3)_F\times SU(2)_S\times
U(6)_{conf}$. Consequently the baryons with the same radial structure
and the same permutational FS-symmetry but different flavor or (and)
spin symmetries will have different mass. \\

Because of the flavor dependent factor $\vec\lambda_i^F \cdot
\vec\lambda_j^F$ the chiral boson exchange interaction (1.1)
will lead to orderings of the positive and negative parity
states in the baryon spectra, which agree with the observed
ones in all sectors. In the case of the spectrum of the nucleon
the strength of the
chiral interaction
between the constituent quarks is sufficient to shift
the lowest positive parity state
in the $N$=2 band (the $N(1440)$) below the negative parity states
in the $N$=1 band ($N(1520)$, $N(1535)$).
In the spectrum of the $\Lambda$ on the other hand it is the
negative parity flavor singlet states (the $\Lambda(1405)$ and
the $\Lambda(1520)$) that remain the lowest lying resonances,
again in agreement with experiment.
The mass splittings between the baryons with different strangeness
and between the $\Lambda$ and the $\Sigma$
which have identical flavor, spin and flavor-spin
symmetries arise from the explicit
breaking of the $SU(3)_F$ symmetry that is caused by the mass
splitting of the pseudoscalar meson octet and the different
masses of the u,d and the s quarks.\\

In section 2 below we review the role of chiral symmetry in the
quark based models for the baryons and the general justification for
considering the baryons to be
formed of constituent quarks that interact by
exchanging pseudoscalar mesons.
This section also contains a comparison between the chiral
boson exchange interaction model and the commonly used perturbative
gluon exchange interaction model, along with the proof of why the
latter leads to incorrect ordering of positive and negative
parity states in the spectra.
Section 3 contains a
description of
the chiral boson mediated interaction and section 4 a description of
the  algebraic structure of the harmonic
oscillator basis states.  In section 5  the symmetry
properties of the  interaction are described and a baryon mass formula
is derived to first order in the chiral interaction.
In section
6 we describe the spectra of the nucleon, the $\Delta$-resonance
and the $\Lambda$ hyperon as they are predicted with the
$SU(3)_F$ symmetric chiral boson interaction (1.1).
The effect of the $SU(3)_F$ breaking in the interaction
that arises from the the quark and meson mass differences
is considered in section 7, where the mass splitting
of the baryon octet and decuplet states is considered.
This leads to set of
new mass formulas for the octet and decuplet states, as
well as for the corresponding excited states. In this section
we
give numerical results for the spectra of all the strange
hyperons as well as the nucleon and the $\Delta$.
In section 8 we
discuss the role of the tensor force associated to the
pseudoscalar-exchange interaction and in section 9
the exceptionally large splitting in the flavor
singlet $\Lambda$(1405)-$\Lambda(1520)$ doublet.
In section 10 we discuss the role of the exchange current
corrections to the baryon magnetic moments that are
associated with the pseudoscalar exchange interaction.
Section 11
contains a discussion of the framework and implications
of the results and section 12 some general comments on the
quark model basis for the meson exchange description of
the nuclear force, and on the $q\bar q$ interaction and
the meson spectrum. \\

\vspace{1cm}

\centerline{\bf 2. Chiral Symmetry and the Quark Model}

\vspace{1cm}

The importance of the chiral symmetry for strong interactions
was realized early on (for an
early review and references see \cite {Pagels}).
This symmetry, which is almost exact in the light u and d flavor
sector is however
only approximate in QCD when strangeness is included,
because of the large mass of the s-quark.
Nevertheless even in 3-flavor QCD the current quark masses may, in
a first approximation, be set to zero (the chiral limit), and
their deviation from zero treated as a perturbation.
The small finite masses of the current
quark masses are however very important for the (finite) masses of the
 mesons. In the chiral limit all
members of the pseudoscalar octet ($\pi^+,\pi^-,\pi^0,K^+,K^-,K^0,
{\overline {K}}^0, \eta$) would have zero mass, which is most
clearly seen in the Gell-Mann-Oakes-Renner \cite{GNOR} relations
that relate the pseudoscalar meson masses to the quark condensates:

$$ {m_{\pi^0}}^2 = - \frac {1}{f_\pi^2} (m_u^0 <\overline {u}u> +
m_d^0 <\overline {d}d>) + O({m_{u,d}^0}^2),$$

$$ {m_{\pi^{+,-}}}^2 = - \frac {1}{f_\pi^2} \frac {m_u^0+m_d^0}
{2} (<\overline {u}u> + <\overline {d}d>) + O({m_{u,d}^0}^2),$$

$$ {m_{K^{+,-}}}^2 = - \frac {1}{f_\pi^2} \frac {m_u^0+m_s^0}
{2} (<\overline {u}u> + <\overline {s}s>) + O({m_{u,s}^0}^2),$$

$$ {m_{K^0,{\overline {K}}^0}}^2 = - \frac {1}{f_\pi^2} \frac {m_d^0+m_s^0}
{2} (<\overline {d}d> + <\overline {s}s>) + O({m_{d,s}^0}^2), $$

$$ {m_{\eta}}^2 = - \frac {1}{3f_\pi^2} (m_u^0 <\overline {u}u> +
m_d^0 <\overline {d}d> + 4m_s^0 <\overline {s}s>) + O({m_{u,d,s}^0}^2).
\eqno(2.1)$$
Here $<\overline {u}u>$, $<\overline {d}d>$ and $<\overline {s}s>$
are the quark condensates of the QCD vacuum which are approximately
equal in magnitude \cite {Shifman} ($<\overline {q}q> \simeq -
(240-250 MeV)^3$).
The nonzero values of the quark condensates, which represent the
order parameter,
is direct consequence of (and evidence for) the spontaneously broken
chiral symmetry in the QCD vacuum. Thus all the
pseudoscalar mesons above are
approximate Goldstone bosons, the nonzero masses of which are determined by the
corresponding current quark masses. The masses and structure of the baryons at
low energy is quite in contrast mainly
determined by the spontaneous
breaking of the chiral symmetry and the effective confining interaction,
and hence only weakly depend on
the current quark masses. The role of the current quark masses
in the structure of the baryons is only to break the $SU(3)_F$
symmetry in the baryon spectrum.\\

The importance of the constraints posed by chiral symmetry
for the quark bag models for the baryons \cite{Chodos} was recognized
soon after its development.
The bag surface term, $\sim \overline {\Psi} \Psi
\delta (R-r)$,  breaks the chiral
symmetry and requires introduction of a compensating chiral meson
field, which couples to the massless quarks on the surface of the bag with
\cite{Chthorn} or without \cite{Brown} this field existing inside of a bag.
An alternative to this surface coupled
version is the volume coupled version \cite{Thomas}. In the early
bag models
with restored chiral symmetry the
massless current quarks within the bag
were assumed to interact not only by
perturbative gluon exchange but also through meson exchange.
Baryon and meson masses and other
static properties have been derived in such
bag models with pion and gluon
exchange interactions e.g. in
refs.
\cite{Jaf,MyBr,The,Mul,Fi,Umino}.
\\

If the sharp surface confinement is replaced by a linear scalar
confining interaction
one obtains a chiral potential model, in which
the chiral field is coupled to the massless quarks moving in
the confining potential \cite{Weise}.  Robson has shown
that
good predictions for the
splittings in the baryon octet-decuplet as well as some
of the excitations in the nucleon spectrum can be achieved
with this picture if the
quarks are assumed to interact by pseudoscalar meson exchange alone
without
gluon exchange \cite{ROB}.
In these models
the chiral field only has the character of a compensating field
rather than a collective low frequency Goldstone quark-antiquark
excitation (the possibility of a nonzero quark condensate
was not addressed). A general limitation of all bag and bag-like models is
of course the lack of translational
invariance, which is important for a realistic description of the
excited states.\\

Common to these models is that the breaking of chiral symmetry
arises from the confining interaction.
This point of view contrasts with that of Manohar and Georgi
\cite{MAG}, who pointed out that there should
be two different scales in QCD, with 3 flavors. At the first one
of these,
$\Lambda_{\chi SB} \simeq 4 \pi f_\pi \simeq$ 1 GeV, the
spontaneous breaking of the chiral symmetry occurs, and hence
at distances
beyond $ \frac {1}{\Lambda_{\chi SB}} \simeq$ 0.2 fm the valence
current quarks acquire their dynamical (constituent) mass
(called "chiral quarks" in \cite{MAG}) and the Goldstone bosons
(mesons) appear. The other scale, $\Lambda_{QCD}
\simeq 100-300$ MeV,
is that which characterizes confinement, and the inverse of this
scale roughly coincides with the linear size of
a baryon. Between these two scales then the effective Lagrangian
should be formed out of the gluon fields that provide
a confining mechanism as well as of the
constituent quark and pseudoscalar meson fields. Manohar and Georgi
did not, however, specify whether the baryons should be
descibed as bound qqq states or as chiral solitons.\\

The chiral symmetry breaking scale above fits well with that which
appears in the instanton liquid picture of the QCD vacuum
\cite{SHU,DIA}. In this model the quark condensates (i.e. equilibrium
of virtual quark-antiquark pairs in the vacuum state)
as well as the gluon condensate
are supported by instanton fluctuations of a size $\sim 0.3$ fm.
The instanton liquid picture for QCD vacuum is confirmed by recent
lattice QCD calculations \cite{Negele} which show
that selective removal of all configurations of the gluon field
except for the instantons does not change the vacuum correlation
functions of hadronic currents \cite{Shu1} and density-density
correlation functions in hadronic bound state.
Dyakonov and Petrov \cite {DIA} suggested that at low momenta
(i.e. beyond the chiral symmetry breaking scale) QCD should
be approximated by an effective chiral Lagrangian of the sigma-model
type that contains valence quarks with dynamical (constituent)
masses and meson fields. They considered
a nucleon as three constituent quarks moving independently
of one another
in a self-consistent chiral field of the hedgehog
form \cite{DIAP}. In this picture the excited baryon states appear
as rotational
excitations and no explicit confining interaction is included.
A very similar description for the nucleon was suggested within so called
"chiral quark models" \cite{KAH,BIR}.\\

The spontaneous breaking of chiral symmetry  and its consequences
- the dynamical quark mass generation, the appearance of the
quark condensate and pseudoscalar mesons as Goldstone
excitations are well illustrated by the Nambu and Jona-Lasinio
model \cite{Nambu,Vogl}.
This model lacks a confining
interaction, which as argued below is essential for a realistic
description of the properties of the baryon physics.\\

The chiral field interaction (1.1) between the constituent
quarks should be contrasted in form with the
color-magnetic interaction \cite {DERU}

$$H_c\sim-\alpha_s\sum_{i<j} \frac {\pi}{6m_im_j} \vec \lambda_i^C\cdot
\vec \lambda_j^C \vec
\sigma_i\cdot \vec \sigma_j\delta(\vec r_{ij}),\eqno(2.2)$$

\noindent
where the $\{\vec \lambda_i^C\}$:s are color $SU(3)$ matrices, and
which should
be important in the region of explicit chiral symmetry at short
distances and high energy. It is in fact this color-magnetic
interaction, which has been used in earlier attempts to describe the baryon
spectra with the constituent quark model
\cite{IGK1,IGK2,CAI}. Although many of the
qualitative and some of the quantitative features of the
fine structure of the baryon
spectra can be described by the interaction (2.2) a number
of outstanding features have proven hard to explain in this approach.\\

The most obvious one of these is the different ordering of the
 positive and negative parity resonances in the
spectra of
the nucleon and delta
on the one hand and the $\Lambda$
hyperon on the other, and in particular the difficulty in describing the
low masses of the
$\Lambda(1405)$, $N(1440)$, $\Delta(1600)$ and $\Sigma(1660)$
 resonances.
A second such feature
is the absence of empirical
indications
for the large spin-orbit interaction
that should accompany the
color-magnetic interaction (2.2) \cite{DERU}.
Although it has been suggested that the latter problem could be
overcome by decreasing of the coupling constant
due to smearing of the contact interaction (2.2) and
by a partial cancellation against the spin-orbit
interaction that is associated with Thomas precession \cite{CAI},
the first problem is inherent to the color operator structure
$ \vec \lambda_i^C\cdot
\vec \lambda_j^C$ of the one gluon exchange interaction
and to the antisymmetry of the color part of the baryon wave function
and cannot be overcome by changing of the radial behavior of
the contact interaction (2.2) and of the confining potential.
Indeed the interaction (2.2) is attractive in color-spin
symmetric quark pair states and repulsive in antisymmetric ones:

$$<[f_{ij}]_C\times [f_{ij}]_S|\vec \lambda^C_i \cdot \vec \lambda_j^C
\vec\sigma_i \cdot \vec \sigma_j|[f_{ij}]_C \times [f_{ij}]_S>$$
$$=\left\{\begin{array}{rr}  8 &
[11]_C,[11]_S:[2]_{CS} \\  -{8\over
3} & [11]_C,[2]_S:[11]_{CS}\end{array}\right..\eqno(2.3)$$

As a consequence of this and
the confining interaction the $\frac{1}{2}^+$
$N(1440)$, which belongs to the N=2 band, should have higher mass
than the $\frac{1}{2}^-$ $N(1535)$ (N=1) as both
have the same mixed color-spin symmetry. Similarly the
$\frac{3}{2}^+$ $\Delta(1600)$ (again N=2), the color-spin state
of which is totally antisymmetric, should have substantially
larger mass than the mixed CS-symmetry N=1 state $\frac{3}{2}^-$
$\Delta(1700)$. Both of these predictions are in conflict
with experiment.\\

We shall show below that the chiral
pseudoscalar interaction (1.1) provides a simpler description of the
fine structure of the
 baryon spectra at the excitation energy up to 1 GeV,
that automatically implies the reversal of the
ordering of the even and odd parity states between the nucleon and
$\Lambda$ hyperon spectra. Moreover we show that when one
matrix element of the effective interaction potential $V(r)$
in (1.1) for each oscillator shell is extracted from the empirical
mass splittings
a quite satisfactory description of the fine structure
of the whole low lying baryon spectrum is
achieved already in lowest order. Finally the overall small
spin-orbit splitting in the baryon spectrum
is qualitatively explained by the absence of any
spin-orbit component in the pseudoscalar
exchange interaction.
This then suggests that it is the chiral field interaction (1.1),
which plays the dominant role in ordering the baryon spectrum in
the region of hidden chiral symmetry, and that the
perturbative gluon exchange interaction becomes
important only at length scales smaller than that of the
spontaneous chiral symmetry breaking where no constituent quarks.\\

In addition to the indication against strong gluon exchange
interactions at low energy that is provided by the
cooled  QCD lattice calculations of Chu et al. \cite{Negele}
mentioned above, there is good evidence
from the recent lattice QCD calculations by Liu and Dong \cite{Liu}
that the splittings  $N - \Delta$ and $\pi - \rho$ are not due to
the one-gluon exchange interaction between quarks. To show this
Liu and Dong measured these mass splittings with two approximations.
The first one is the standard quenched approximation, which neglects
sea quark closed loop diagrams generated by gluon lines.
This quenched approximation contains however part of the antiquark
effects related to the Z graphs formed of valence quark lines
\cite {Cohen}. The second one is so called
"valence approximation" where the quarks are limited to propagating
only forward in time (i.e. no Z-graphs and related quark-antiquark
pairs). The gluon exchange and all other possible gluon
configurations including instantons are exactly the same
within both approximations. As expected finite $N - \Delta$
and $\pi - \rho$ splittings are observed in the quenched approximation,
but not in the valence approximation, in which
the $\Delta$ and the $N$ and the $\rho$ and the $\pi$
become degenerate within
errors \cite{Liu}.
Since the one-gluon exchange is not switched
off in the valence approximation this indicates that the hyperfine splitting
does not arise from the one-gluon exchange interaction.\\

\vspace{1cm}

\centerline{\bf 3. The Chiral Boson exchange interaction}
\vspace{1cm}
In an effective chiral symmetric description of baryon
structure based on the constituent quark model the
coupling of the quarks and the pseudoscalar Goldstone
bosons will (in the $SU(3)_F$ symmetric approximation) have
the form $ig\bar\psi\gamma_5\vec\lambda^F
\cdot \vec\phi\psi$, where $\psi$ is the fermion constituent quark
field operator and $\vec\phi$ the octet boson field
operator, and g is a coupling constant. A coupling of this
form in a nonrelativistic reduction for a constituent quark spinors
will -- to lowest order -- give rise to a Yukawa interaction
between the constituent quarks, the spin-spin component of which has
the form
$$V_Y (r_{ij})= \frac{g^2}{4\pi}\frac{1}{3}\frac{1}{4m_im_j}
\vec\sigma_i\cdot\vec\sigma_j\vec\lambda_i^F\cdot\vec\lambda_j^F
\{\mu^2\frac{e^{-\mu r_{ij}}}{ r_{ij}}-4\pi\delta (\vec r_{ij})\}
.\eqno(3.1)$$
Here $m_i$ and $m_j$ denote masses of the interacting quarks
and $\mu$ that of the meson. There will also be an associated
tensor component, which will be discussed in section 8 below.
Because of the short range of the baryon wavefunctions the role
 of the $\delta$ function term is of crucial importance,
although the latter is expected to be smeared out by the finite
size of the constituent quarks and pseudoscalar mesons.\\

Along with the pseudoscalar bosons chiral symmetry requires an
accompanying scalar meson field $\phi_\sigma$ to complete the chiral
multiplet. This also contributes an effective interaction between the
constituent quarks. As the main component of this scalar meson
exchange interaction is a spin- and flavor-independent attractive
interaction it contributes to the effective confining interaction, but
not to the fine structure of the spectrum except through the
associated weak spin-orbit interaction, which will be discussed in
section 9 below. \\

At short range the simple form (3.1) of the chiral boson exchange
interaction cannot be expected to be realistic, and should only
be taken to be suggestive.
Because of the finite spatial extent of both the constituent
quarks and the pseudoscalar mesons
that the delta function in (3.1) should be replaced by a finite
function, with a range of 0.6-0.7 fm as suggested
by the
spatial extent of the mesons.
In addition the radial behaviour of the Yukawa
potential (3.1) is valid only if the boson field
satisfies linear Klein-Gordon equation. The chiral symmetry
requirements for the effective Lagrangian, which contains
constituent quarks as well as boson fields
imply that these boson fields cannot be described by linear
equations near their source. Therefore it is only at large
distances where the amplitude of the boson fields is small that
the quark-quark interaction reduces to the simple Yukawa form.
At this stage the proper procedure should be to avoid further specific
assumptions about the short range behavior of
$V(r)$ in
(1.1) and instead to extract the required matrix elements of it
from the baryon spectrum and to reconstruct by this an approximate
radial form of $V(r)$.
The overall -- sign in the
effective chiral boson interaction in (1.1) corresponds to that of this
short range term in the Yukawa interaction.\\

The flavor structure of the pseudoscalar octet exchange interaction
in (1.1) between two quarks i and j should be understood as
follows

$$V(r_{ij}) \vec {\lambda^F_i} \cdot \vec {\lambda^F_j} =
\sum_{a=1}^3 V_{\pi}(r_{ij}) \lambda_i^a \lambda_j^a
+\sum_{a=4}^7 V_K(r_{ij}) \lambda_i^a \lambda_j^a
+V_{\eta}(r_{ij}) \lambda_i^8 \lambda_j^8. \eqno(3.2)$$
The first term in (3.2) represents the pion-exchange interaction,
which acts only between
light quarks. The second term represents the kaon
exchange interaction,
which takes place in u-s and d-s pair states. The $\eta$-
exchange, which is represented by the third term, is allowed
in all quark pair states. In the $SU(3)_F$ symmetric limit
the constituent quark masses would be equal ($m_u = m_d = m_s$),
the pseudoscalar octet would be degenerate and the meson-constituent
quark coupling constant
would be flavor independent. In this limit the
form of the pseudoscalar exchange interaction reduces to (1.1),
which does not break the $SU(3)_F$ invariance of the baryon
spectrum. Beyond this limit the pion, kaon and $\eta$
exchange interactions will differ ($V_\pi \not= V_K \not= V_\eta$)
because of the difference between the strange and u, d quark
constituent masses ($m_{u,d} \not= m_s$), and because of the
mass splitting within the pseudoscalar octet
($\mu_\pi \not= \mu_K \not= \mu_\eta$) (and possibly also because
of flavor dependence in the meson-quark coupling constant).
As pion exchange and
kaon exchange takes place only for quark pairs
of unique mass (we neglect the possible small mass difference between
u and d constituent quarks)
the dependence on the quark mass can be
absorbed into the corresponding
potential functions $V_\pi$ and $V_K$. As on the other hand
$\eta$ exchange is
possible in all light and strange
quark pair combinations
with different mass the
potential function $V_\eta$ should be expected to be
flavor dependent as it is denoted below by the
corresponding subscripts. The source of both the $SU(3)_F$ symmetry
breaking constituent quark mass differences and the $SU(3)_F$ symmetry
breaking mass splitting of the pseudoscalar octet is
the explicit chiral symmetry breaking in QCD.\\

The flavor matrix elements of the interaction are\\

$$ <~[f_{ij}]_F ~T_{ij}~\mid  \vec{\sigma_i} \cdot \vec{\sigma_j}
\sum_{a=1}^8 V^a(r_{ij}) \lambda_i^a \lambda_j^a
{}~\mid ~[f_{ij}]_F~ T_{ij}~> $$

$$~~=~~\vec{\sigma_i} \cdot \vec{\sigma_j} \times \left\{ \begin{array}{lll}
V_\pi + \frac {1} {3} V_\eta^{uu}, & {\rm if} \; [2]_F, & T_{ij}=1 \\
2V_K - \frac {2} {3} V_\eta^{us}, & {\rm if} \; [2]_F, & T_{ij}= \frac {1}{2}\\
\frac {4} {3} V_\eta^{ss}, & {\rm if} \; [2]_F, & T_{ij}=0 \\
-2V_K - \frac {2}{3} V_\eta^{us}, & {\rm if} \; [11]_F, & T_{ij}=\frac {1}{2}\\
-3V_\pi + \frac {1} {3} V_\eta^{uu}, & {\rm if} \; [11]_F, & T_{ij}=0
\end{array} \!\!\right.
.\eqno (3.3)$$

\noindent
Here the Young pattern $[f_{ij}]_F$ denotes the flavor permutational
symmetry in the quark pair i, j (the symbol [2] represents
the Young pattern with two boxes in
in first row and [11] that with two boxes in one column).
The total isospin of the pair state is denoted $T_{ij}$. The subscripts
uu, us, ss on the $\eta$
exchange
potential $V_\eta$ indicate that the potential acts in
pairs of two light, one light and one strange and
two strange quarks respectively.

\vspace{1cm}

\centerline{\bf 4. Algebraic Structure of the Oscillator Wavefunctions}
\vspace{1cm}
The confining interaction between two constituent quarks i, j
will be taken to have the harmonic oscillator form

$$V_{conf}(\vec r_{ij})=V_0+\frac{1}{6} m\omega^2
(\vec r_i - \vec r_j)^2,\eqno(4.1)$$

\noindent
where $m$ is the mass of the constituent quark  and
$\omega$ is the angular frequence of the oscillator
interaction. For simplicity we neglect here the mass
difference between the light and strange constituent quarks. The
Hamiltonian for the unperturbed basis states of the
3 quark system then takes the form

$$H_0=\sum_{i=1}^3\frac{\vec p_i^2}{2 m_i}
-\frac{\vec P_{cm}^2}{6m}+
\frac{1}{6}\sum_{i<j}m\omega^2(\vec r_i-\vec r_j)^2
+3 V_0.\eqno(4.2)$$

\noindent
Here $\vec P_{cm}$ denotes the total momentum of the baryon.
 The exact eigenvalues
of this Hamiltonian are
$$E_0=3V_0+(N+3)\hbar\omega,\eqno(4.3)$$
where $N$ is the number of excitation quanta in the state.\\

The overall
orbital symmetry of A particles interacting each other through
harmonic forces is $U(3(A-1))$ which
in the present case reduces to $U(6)$.
The eigenvalue (4.3) is highly degenerate and therefore
additional quantum numbers are required to characterize
an eigenstate uniquely.\\

The spatial part of the three body wave function is determined by
the following quantum numbers:
$$\vert N (\lambda \mu) L [f]_X (r)_X >. \eqno(4.4)$$
Here we use the notations of the translationally invariant
shell model (TISM) \cite{Kurd}. The wave functions are
exact solutions of the three-body
Schr\"odinger equation with the Hamiltonian (4.2) and coincide
with the corresponding harmonic oscillator shell
model states after removal of the center-of-mass motion
from the latter. The Elliott symbol $(\lambda \mu)$ determines
the SU(3) harmonic oscillator multiplet with the dimension
$dim(\lambda \mu) = \frac {1} {2} (\lambda +1)(\mu + 1)
(\lambda + \mu + 2)$ and $L$ is the total orbital angular momentum.
The allowed values of $L$ that are
compatible with a given Elliott symbol
$(\lambda \mu)$ are given by the Elliott formula
\cite{Kurd,Elliott}. \\

The spatial permutational symmetry of the state
is indicated by the Young pattern (diagram) $[f]_X$, where f is
a sequence of integers that indicate the number
of boxes in the successive rows of the corresponding Young patterns.
Thus $[3]$ represents the completely symmetric state, $[111]$
the  completely
antisymmetric one and $[21]$ states of mixed symmetry. Finally,
$(r)_X$ is the Yamanouchi symbol which determines the basis
vector of the irreducible representation $[f]_X$ of the permutation
group $S_3$. The Yamanouchi symbol is uniquely connected with
the standard Young tableau (i.e. with the Young pattern where the numbers
of particles are put in boxes in regular sequence). For the totally
symmetrical $[3]$ or totally antisymmetrical $[111]$ representations
they are unique - $(111)$ and $(123)$, respectively. For the mixed
symmetry state $[21]$ there are two different basis vectors
determined by the Yamanouchi symbols $(112)$ (i.e.
the first and the second particles are in the first row and the
third particle is in the second row of the Young pattern) and
$(121)$ (the first and the third particles are in the first row
and the second particle is in the second row). All the necessary
functions (4.4) are well known and can be found e.g. in \cite {GLOK}.\\

The Hamiltonian (4.2) does not depend on the spin and flavor degrees of
freedom. Thus to provide the full set of zero order wave functions
one should construct all possible color-flavor-spin parts that
are compatible
with a given spatial wave function. The color part of the
wave function is totally antisymmetric ($[111]_C$) and
therefore the Pauli principle requires
the spatial-flavor-spin part should
be symmetric:

$$ [111]_{CXFS} = [111]_C \times [3]_{XFS}. \eqno(4.5)$$

\noindent
The color part of the wave function can be factored out
and it will be suppressed
in the expressions below as the interaction (1.1) is independent
of color. A possible color dependence of the confining interaction
of the form
$\vec \lambda^C_i \cdot \vec \lambda_j^C$ is inessential for baryon
states as the corresponding matrix element

$$<[11]_C|\vec \lambda^C_i \cdot \vec \lambda_j^C
|[11]_C > = -{8 \over3}\eqno(4.6)$$

\noindent
is the same for all quark pair states and hence can
be absorbed into definition
of the effective confining interaction.\\

The total symmetry of the spatial-flavor-spin wave function
implies that the permutational symmetry $[f]_X$ of the orbital part
(4.4) and the permutational symmetry $[f]_{FS}$ of the flavor-spin
part have to coincide: $[f]_X = [f]_{FS}$. By the general
rules the symmetrical spatial-flavor-spin wave functions should be
constructed as

$$\mid N (\lambda \mu) L [f]_X [f]_{FS} [f]_F [f]_S Y T>$$

$$= \frac {1} {\sqrt{dim [f]_X}} \sum_{(r)_X=(r)_{FS}}
\mid N (\lambda \mu) L [f]_X (r)_X>~
\mid [f]_{FS} [f]_F [f]_S Y T (r)_{FS}>, \eqno(4.7)$$

\noindent
where $[f]_F$ and $[f]_S$ denote permutational flavor and
spin symmetries, respectively and $Y$ is hypercharge and $T$
is isospin of a baryon. Obviously, $dim [111] = dim [3] = 1$,
$dim [21] = 2$. Note that $[f]_S$
uniquely determines the total spin $S$ as one half of the difference
of the first and second rows in the spin Young pattern above.
It is also understood that the orbital momentum $L$ and spin $S$
are coupled to the total angular momentum $J$.\\

The flavor-spin part of the 3 quark states in (4.7) is
easily constructed by the fractional parentage
expansion for the separation of one particle

$$|[f]_{FS}[f]_F[f]_S YT(r)_{FS};ts>=
\sum_{T_{12}t_{12}T_3t_3S_{12}s_{12}S_3s_3} \Gamma$$

$$\times ~ |[f_{12}]_{FS}[f_{12}]_F  Y_{12}T_{12}S_{12};t_{12}s_{12}>
{}~|Y_3 T_3 S_3=\frac{1}{2};t_{3}s_3>$$
$$\times ~ (T_{12} t_{12} T_3 t_{3}|Tt)(S_{12} s_{12} S_3 s_3|Ss).\eqno(4.8)$$
Here
$[f_{12}]$  denotes Young patterns for the symmetries
of the two-particle states. Obviously the flavor-spin
symmetry $[f_{12}]_{FS}$
is uniquely determined by the Yamanouchi symbol $(r)_{FS}$.
The quantum numbers for the
hypercharge, isospin and spin of the particle pair 12 and the
single particle 3 are indicated by subscripts. Finally the
fractional parentage coefficient $\Gamma$ can be presented
as a product of two factors \cite{fpc}. The first is the scalar
factor of the Clebsch-Gordan coefficient for the group
$SU(6)_{FS}$ in the reduction
$SU(6)_{FS} \supset SU(3)_F \times SU(2)_S$,
which is determined only by invariants of the groups $SU(6)_{FS}$,
$SU(3)_F$ and $SU(2)_S$ and does not depend on isospin, hypercharge
nor on the third components of isospin and spin:
$$\left([f_{12}]_{FS}~[f_{12}]_F~ S_{12} ; ~[1]_{FS}~ [1]
_F~ S_3=\frac{1}{2}~||~[f]_{FS}~[f]_F S\right).\eqno(4.9a)$$

\noindent
The second is the isoscalar factor of the $SU(3)_F$ Clebsch-Gordan
coefficient:
$$\left([f_{12}]_F~Y_{12}~T_{12};~ [1]_F~Y_3 ~T_3 ~||~
[f]_F~Y~T\right).\eqno(4.9b)$$
The  coefficients (4.9a) are listed in Table 1 and the
(4.9b) ones in Tables 2a and 2b.\\

In the following we shall use the quantum numbers above
to characterize wave function:

$$ \Psi = \mid N (\lambda \mu) L [f]_X [f]_{FS}[f]_F [f]_S YT>. \eqno(4.10)$$

\noindent
The explicit indication of the permutational
symmetry in the notation for the states is more convenient than the
indication of dimension of the corresponding multiplets, which
is conventional in baryon and meson
spectroscopy. This is because the symmetry properties of
the interaction (1.1) (see next Section) together with the permutational
symmetries $[f]_{FS}$, $[f]_F$ and $[f]_S$ makes it
very transparent for which states the interaction (1.1)
is most  attractive, and hence to understand
the ordering of the states in the spectrum. All the required
dimensions are easily calculated for the given Young patterns
according to the general rules \cite{Ham} and are listed below:

$$SU(2)_S :\left\{\begin{array}{lll}  4, \quad [3]_S \\
 2, \quad [21]_S \end{array}\right.,\eqno(4.11)$$

$$SU(3)_F :\left\{\begin{array}{lll}  10,  \quad
 [3]_F \\  8,  \quad [21]_F\\ 1,  \quad [111]_F
\end{array}\right.,\eqno(4.12)$$

$$SU(6)_{FS} :\left\{\begin{array}{lll}  56,  \quad
 [3]_{FS} \\  70,  \quad [21]_{FS}\\ 20,  \quad [111]_{FS}
\end{array}\right..\eqno(4.13)$$\\
The symmetry structure of the zero order wave functions
is $SU(6)_{FS} \times U(6)_{conf}$. Thus, when the only
interaction between the quarks is the flavor- and spin-independent
harmonic confining interaction and all quarks have equal mass
the baryon spectrum would be organized in multiplets of the
group above. In this case the baryon masses would be determined
solely by the orbital structure and by the constituent quark mass
and the spectrum would be organized in an alternating sequence
of positive and negative parity states: the ground states (N=0,
positive parity), the first excited band (N=1, negative parity),
the second excited band (N=2, positive parity) and so on.\\

If the confining potential is not harmonic,
but some other possible monotonically increasing central potential,
the symmetry structure of the zero order wave functions reduces
to $SU(6)_{FS} \times O(3)$.

\vspace{1cm}

\centerline{\bf 5. The Symmetry Structure of the Chiral Boson
Exchange Interaction}
\centerline {\bf and the Baryon Mass Formula}
\vspace{1cm}

In the $SU(3)_F$ limit all pseudoscalar octet mesons would be
degenerate and also $m_u = m_d = m_s$. In this limit
$V_\pi = V_K = V_\eta = V$ in (3.2) and (3.3). The flavor-spin
two-quark matrix elements of the
the chiral boson exchange interaction
are in this case:

$$<[f_{ij}]_F\times [f_{ij}]_S : [f_{ij}]_{FS}
{}~| -V(r_{ij})\vec \lambda^F_i \cdot \vec \lambda_j^F
\vec\sigma_i \cdot \vec \sigma_j
{}~|~[f_{ij}]_F \times [f_{ij}]_S : [f_{ij}]_{FS}>$$
$$=\left\{\begin{array}{rr} -{4\over 3}V(r_{ij})& [2]_F,[2]_S:[2]_{FS} \\
-8V(r_{ij}) & [11]_F,[11]_S:[2]_{FS} \\
4V(r_{ij}) & [2]_F,[11]_S:[11]_{FS}\\ {8\over
3}V(r_{ij}) & [11]_F,[2]_S:[11]_{FS}\end{array}\right..\eqno(5.1)$$

\noindent
{}From these the following important properties may be inferred:

(i) At short range where $V(r_{ij})$ is positive the chiral
interaction (1.1) is attractive in the symmetrical FS pairs and
repulsive in the antisymmetrical ones. At large distances the potential
function $V(r_{ij})$ becomes negative and the situation is
reversed.

(ii) At short range  among the $FS$-symmetrical pairs
the flavor antisymmetrical pairs experience
a much larger attractive interaction than the flavor-symmetrical
ones and among the FS-antisymmetrical pairs
the strength of the repulsion in flavor-antisymmetrical
pairs is considerably weaker than in symmetrical ones.\\

Given these properties we conclude that
with the given flavor symmetry the more symmetrical flavor-
spin Young pattern for a baryon - the more attractive contribution at short
range comes from the interaction (1.1). With two identical
flavor-spin Young patterns $[f]_{FS}$ the attractive contribution
at short range is larger in the case with the more antisymmetrical
flavor Young pattern $[f]_F$.\\

\bigskip
When the
boson exchange interaction is treated in first order perturbation
theory the mass of the baryon states takes the form

$$M=M_0+N\hbar\omega+ \delta M_\chi, \eqno(5.2)$$

\noindent
where the chiral interaction contribution is

$$ \delta M_\chi = <\Psi|H_\chi|\Psi>, \eqno(5.3)$$

\noindent
and

$$M_0 = \sum_{i=1}^3 {m_i} + 3(V_0 + \hbar \omega). \eqno(5.4)$$

The interaction (1.1) is diagonal in states of definite
orbital angular momentum $L$ and good $[f]_{FS}, [f]_F, [f]_S$
symmetries and thus there is no configuration mixing in
zero order perturbation theory of the states that are degenerate
in energy at lowest order. The associated tensor interaction,
expected to be weak as mentioned above, does however mix states
with equal $J$ and flavor symmetry.\\

Due to the overall antisymmetry of the wavefunctions $\Psi$,
the chiral contribution (5.3) can be expressed as

$$ <\Psi|H_\chi|\Psi> = 3<\Psi| -V(r_{12})\vec \lambda^F_1
\cdot \vec \lambda_2^F
\vec\sigma_1 \cdot \vec \sigma_2 ~|\Psi>. \eqno(5.5)$$

\noindent
This is readily evaluated with the explicit expression
for the wave function (4.7) and the matrix elements (3.3) for
the flavor part.
The result can be found in Tables 3-10 below and
is a linear combination of the spatial matrix elements of
the two-body potential $V(r_{12})$, defined as

$$P_{nl}^k=<\varphi_{nlm}(\vec r_{12})
|V_k(r_{12})|\varphi_{nlm}(\vec r_{12})>.\eqno (5.6)$$
Here $\varphi_{nlm}(\vec r_{12})$
represents the oscillator wavefunction
with n excited quanta, and $k$ the exchanged meson.
 As we shall only consider
the baryon states in the $N\le 2$ bands we shall only need
the 4 radial matrix elements $P_{00},P_{11},P_{20}$ and
$P_{22}$ for the numerical construction of the spectrum.\\

\newpage
\centerline{\bf 6. The Structure of the Baryon Spectrum}

\vspace{1cm}

Consider first for the purposes of illustration a schematic model
which neglects the radial dependence
of the potential function $V(r)$ in (1.1).
In this model all the radial integrals $P_{nl}^{k}$
(5.6) will have the same constant value $C_\chi$.\\

The 3-quark states in the baryon spectrum have
the following flavor-spin symmetries:
$$[3]_{FS}[21]_F[21]_S,\quad [3]_{FS}[3]_F[3]_S,\quad
[21]_{FS}[21]_F[21]_S,$$
$$[21]_{FS}[3]_F[21]_S,\quad
[21]_{FS}[21]_F[3]_S,\quad [21]_{FS}[111]_F[21]_S.\eqno(6.1)$$
For these states the matrix elements (5.5) are
$-14C_\chi$, $-4C_\chi$, $-2C_\chi$, $4C_\chi$, $2C_\chi$ and $-8C_\chi$
 respectively. The first one of these
describes the $S=\frac{1}{2}$ baryons $N,\Lambda,\Sigma$
and $\Xi$ in the ground state band and the second
the corresponding $S=\frac{3}{2}$ resonances $\Delta(1232)$,
$\Sigma(1385)$, $\Xi(1530)$ and $\Omega$. The third describes the lowest
negative parity doublet in all sectors, except for that
of the $\Lambda$, in which case the lowest negative
parity doublet $\Lambda(1405)-\Lambda(1520)$ is a flavor
singlet described by the last one of the flavor-spin states
above.\\

These  matrix elements alone suffice to prove that
the ordering of the positive and negative parity states
in the baryon spectrum will be correctly predicted by
the chiral boson exchange interaction (1.1).
The constant $C_\chi$ may be determined from the
$N-\Delta$ splitting to be 29.3 MeV.
When the radial structure of the interaction (1.1)
is neglected as above the oscillator
parameter $\hbar\omega$
may be determined by the mass differences between the
first excited
$\frac{1}{2}^+$ states and the ground states of the baryons,
which have the same flavor-spin, flavor and spin symmetries
(e.g. $N(1440)-N$, $\Lambda(1600) - \Lambda$, $\Sigma(1660) - \Sigma$), to be
$\hbar\omega \simeq 250$ MeV.
In the $N$ sectors the mass
difference between the lowest
excited ${1\over 2}^+$ ($N(1440)$)
and ${1\over 2}^-$
states ($N(1535)$) will then be
$$N:\quad m({1\over 2}^+)-m({1\over 2}^-)=250\, {\rm
MeV}-C_\chi(14-2)=-102\, {\rm MeV},\eqno(6.3)$$
whereas it for the $\Lambda$ system ($\Lambda(1600),
\Lambda(1405)$) should be

$$\Lambda:\quad m({1\over 2}^+)-m({1\over 2}^-)=250\, {\rm
MeV}-C_\chi(14-8)=74\, {\rm MeV}. \eqno(6.4)$$
This simple example shows how the chiral interaction (1.1)
provides different ordering of the lowest positive and negative excited
states in the spectra of the nucleon and
the $\Lambda$-hyperon. This is a direct
consequence of the symmetry properties of the boson-exchange interaction
discussed in the previous section. That the $SU(2)_T\times SU(2)_S$
version of the interaction (1.1) may be important for the
downshift of the Roper resonance
has in fact been noted earlier \cite{Ob}.\\

Consider now in addition the radial dependence of the potential
with the $SU(3)_F$ invariant version (1.1) of
the chiral boson exchange interaction (i.e. $V_\pi (r)
=V_K (r)=V_\eta(r)$).
The contribution
to all nucleon, $\Delta$ and $\Lambda$ hyperon states from the
boson exchange interaction in terms of the
matrix elements $P_{nl}$ (5.6) are listed in Tables 3 and
4. In this approximate version of the chiral boson
exchange interaction the $\Lambda-N$ and the $\Xi - \Sigma$
mass differences
would solely be ascribed the  mass difference
between the s and u,d quarks since all these baryons have identical
orbital structure and permutational symmetries
and the states in the $\Lambda$-spectrum would be degenerate
with the corresponding states in the $\Sigma$-spectrum which have
equal symmetries.\\

The  oscillator parameter $\hbar\omega$ and the 4 integrals
that appear in the two tables are extracted from
the mass differences between the nucleon and the $\Delta(1232)$,
the $\Delta(1600)$ and
the $N(1440)$, as well as the splittings between the nucleon
and the average mass of the
two pairs of states $N(1535)-N(1520)$ and
$N(1720)-N(1680)$.
This procedure yields the parameter values
$\hbar\omega$=157.4 MeV,
$P_{00}$=29.3 MeV, $P_{11}$=45.2 MeV, $P_{20}$=2.7 MeV and
$P_{22}$=--34.7 MeV. Given these values all other excitation energies
(i.e. differences between the masses of given resonances and
the corresponding ground states)
of the nucleon, $\Delta$- and $\Lambda$-hyperon spectra are
predicted to within $\sim$ 15\% of the empirical values
where known, and well within the uncertainty limits
of those values.
These matrix elements provide a quantitatively satisfactory
description of the $\Lambda$-spectrum
even though they are extracted from the $N-\Delta$ spectrum.
The parameter values above should
be
allowed a considerable uncertainty range in view of
the uncertainty in the empirical values for the
resonance energies. To illustrate this we note that the
description of the resonance energies does not notably
deteriorate if instead the following set of parameter
values were used: $\hbar\omega=$ 227.4 MeV,
$P_{11}=$ 31.2 MeV, $P_{20}$=22.7 MeV and $P_{22}=$--14.4 MeV,
with the same value for $P_{00}$ as above. These latter
values are obtained by taking the $\Delta$(1600) to have
an energy of 1700 MeV. \\

As mentioned above the
symmetrical $FS$ pair states experience an attractive interaction
at short range, whereas antisymmetrical ones experience repulsion.
This
explains why the $[3]_{FS}$ state in the $N(1440)$, $\Delta(1600)$
and
$\Sigma(1660)$ positive parity resonances feels a stronger
attractive interaction than the mixed symmetry state $[21]_{FS}$ in the
$N(1535)$,
$\Delta(1700)$
and $\sum(1750)$ resonances. Consequently the masses of the
positive parity states $N(1440)$, $\Delta(1600)$  and
$\Sigma(1660)$ are shifted
down relative to the other ones, which explains the reversal of
the otherwise expected "normal ordering".
The situation is different in the case of the $\Lambda(1405)$ and
$\Lambda(1600)$, as the flavor state of the $\Lambda(1405)$ is
totally antisymmetric. Because of this the
$\Lambda(1405)$ gains an
attractive energy, which is
comparable to that of the $\Lambda(1600)$, and thus the ordering
suggested by the confining oscillator interaction is maintained.\\

The predicted nucleon (and $\Delta$) spectrum,
which in Table 3 is listed up to $N=2$, contains two groups of
nonconfirmed and
unobserved states. These all belong to the
$N=2$-band. The lowest group is the 4 $\Delta$
states around 1675 MeV, one of which plausibly
corresponds to the 1-star $\Delta(1750)$. The predicted
${3\over 2}^+$ and ${5\over 2}^+$ resonances around 1909 MeV
plausibly correspond to the 1-- and 2--star resonances
$N(1900)$ and $N(2000)$ respectively. The predicted ${1\over 2}^+$
state at 1850 MeV corresponds well with the recent evidence
in favor of a fourth $P_{11}$ state in the 1750 MeV - 1885 MeV
region \cite{MANL,NEFK}. Similarly the predicted
${3\over 2}^+$ state at 1813 MeV corresponds well with the
suggestion of a $P_{13}$ state at 1885 MeV in ref.\cite{MANL}.
The predicted $\Lambda$
spectrum contains one unobserved state in the $N=1$ band and 8 in the
$N=2$ band. As these are predicted to lie close to observed states
with large widths their existence is not ruled out.
The structure of the spectra of the $\Sigma$ and $\Xi$
hyperons are predicted to be similar to that of the
nucleon and the $\Delta$ resonance in Table 3.
However all $\Sigma$ and $\Lambda$ resonances
with equal spatial structure and
with the same
flavor-spin, flavor and spin symmetries are degenerate
within the SU(3)-symmetric version of the boson-exchange
interaction. This degeneracy is lifted by
the $SU(3)_F$ breaking interaction (3.2)-(3.3) that is
treated
in the following section.\\

The relative magnitudes and signs
of the numerical parameter values can be readily understood. If
the potential function $V(\vec r)$ is assumed to have the
form of a Yukawa function with a smeared $\delta$-function
term that is positive  at short range $r\le 0.6-0.7$  fm,
as suggested by the pion size $\sqrt{<r_\pi^2>}=0.66$ fm,
one expects $P_{20}$
to be considerably smaller than $P_{00}$ and $P_{11}$,
as the radial wavefunction for the excited S-state has a node,
and as it extends further into region of where the potential
is negative.
The negative value for $P_{22}$ is also natural as the
corresponding wavefunction is suppressed at short range
and extends well beyond the expected
0 in the potential function.
The relatively small value of the oscillator parameter (157.4 MeV)
leads to the empirical value 0.86 fm for the
nucleon radius $\sqrt{<r^2>}=\sqrt{\hbar/m\omega}$
if the light quark constituent mass is taken to be 330-340 MeV,
as suggested by the magnetic moments of
the nucleon.\\

\vspace{1cm}

\centerline{\bf 7. The  ${\bf SU(3)_F}$ Breaking Chiral Boson
Interaction}

\vspace{1cm}

The contributions to the baryon masses in the $N=0$ band
that arise from the $SU(3)_F$ symmetry breaking version
of the chiral boson interaction (3.2) are listed in Table 5
along with the correction that arises from the difference
$\Delta_q$ between
the masses of the strange and up, down constituent
quarks. In this case we shall distinguish between the different
strengths of the $\eta$-exchange interaction for pairs
of light $(uu)$,
of one light and one strange $(us)$ and two strange $(ss)$ quarks. The
matrix elements to be considered are thus (cf. (5.6)) those of
the pion- and kaon exchange interactions $P_{nl}^{\pi}$ and $P_{nl}^K$
and the $\eta$-exchange interaction matrix elements $P_{nl}^{uu}$,
$P_{nl}^{us}$ and $P_{nl}^{ss}$. As indicated by the Yukawa
interaction (3.1) these matrix elements should be inversely
proportional to the product of the quark masses of the pair state
(this result is common to all $\sigma$-model based interactions). Thus

$$P_{nl}^{us}={m_u\over m_s}P_{nl}^{uu},\quad P_{nl}^{ss}=({m_u\over
m_s})^2P_{nl}^{uu}.\eqno(7.1)$$
Here the usual assumption of equality between the constituent masses
of the up and down quark $(m_u=m_d)$ has been made.\\

To determine the matrix element $P_{00}^{K}$ and $P_{00}^{us}$ we
consider the $\Sigma(1385)-\Sigma$ mass difference, which depends only on
these two integrals:

$$m_{\Sigma(1385)}-m_{\Sigma}=4P_{00}^{us}+6P_{00}^{K} \eqno(7.2)$$
With the assumption that $P_{00}^{us}\simeq P_{00}^{K}$, which is
suggested by the fact that the quark masses are equal in the states,
in which these interactions act, and by the near equality of the kaon
and $\eta$ masses, $\mu_\eta \simeq \mu_K$,
we obtain $P_{00}^{K}=P_{00}^{us}=19.6$ MeV. To determine the integral
$P_{00}^{\pi}$ and the quark mass difference $\Delta_q=m_s-m_u$ we
consider the $N-\Delta$ and $\Lambda-N$ mass splittings:

$$m_\Delta-m_N=12P_{00}^{\pi}-2P_{00}^{uu},\eqno(7.3a)$$
$$m_\Lambda-m_N=6P_{00}^{\pi}-6P_{00}^{K}+\Delta_q.\eqno(7.3b)$$

\noindent
Elimination of $P_{00}^{\pi}$ from these two equations yields
$\Delta_q=121$ MeV if the conventional value 340 MeV is given to
$m_u$. Solving for $P_{00}^{\pi}$ then gives the value $P_{00}^{\pi} =
28.9$ MeV and the quark mass ratio
$m_s/m_u=1.36$. The $\Sigma-\Lambda$ and $\Xi
-\Sigma$ mass differences have the expressions

$$m_\Sigma-m_\Lambda=8P_{00}^{\pi}-4P_{00}^{K}-{4\over
3}P_{00}^{uu}-{8\over 3}P_{00}^{us},\eqno(7.4a)$$

$$m_{\Xi}-m_\Sigma=P_{00}^{\pi}+{1\over 3}P_{00}^{uu}-{4\over
3}P_{00}^{ss}+\Delta_q.\eqno(7.4b)$$
With the matrix element values above these expressions lead to the
values 65 MeV and 139 MeV for these two splittings in good agreement
with the empirical values 77 MeV and 125 MeV respectively.
This explanation
of the octet mass splittings is differs from the early
suggestion for explaining it in terms of an interaction of the form
$\vec\sigma_i\cdot\vec\sigma_j V(r_{ij})/m_i m_j $, with
$V(r_{ij})$ being a flavor independent function
\cite{ZEL,SAK1,SAK2,SAK3}.\\

The predictions for the energies of the
baryon states in the $N=0$ band that are
obtained with the values for the integrals $P_{00}$ above
are listed in Table 5. The predicted values are in remarkably
satisfactory agreement with the empirical values. The largest
deviation occurs for  the $\Omega^-$, the energy of which is
underpredicted by 21 MeV (i.e. by 1\%). The quality of the fit
can be improved by relaxing the requirement
that $\Delta_q = m_s - m_u$ be determined to satisfy
eqs. (7.3) exactly
as above. With the values $\Delta_q =127$ MeV,
$m_u = 340$ MeV (i.e. $m_s/m_u = 1.37$), $P_{00}^\pi = 29.05$ MeV,
$P_{00}^K = 20.1$ MeV
the  deviations between the experimental mass and predicted
mass values are within a few MeV except for the $\Sigma$ and
$\Xi$ where these deviations are about 10 MeV.
 The
numerical values in Table 5 are very similar to those obtained
earlier by Robson
\cite{ROB}, who considered
a similar pseudoscalar meson-exchange model
for the interaction between
 massless current quarks. In Robson's model
the different strengths of the pion, kaon and $\eta$ exchange
interactions at short range were obtained by taking the
 pseudoscalar
meson mediated interactions to be proportional to the inverse square
of the appropriate meson decay constants.\\

	When the matrix elements of the boson exchange
interaction (3.2) and the quark mass difference $\Delta_q$
are eliminated from the expressions for the $N=0$ band
baryons in Table 5 the following mass relations are
obtained:

$$m_\Delta-m_N=m_{\Sigma(1385)}-m_\Sigma
+\frac {3}{2}\left( m_\Sigma-m_\Lambda\right),\eqno(7.5a)$$

$$m_{\Sigma(1385)}-m_\Sigma=m_{\Xi(1530)}-m_\Xi,\eqno(7.5b)$$

$$\frac {1} {3} \left( m_\Omega-m_\Delta\right)
=m_{\Xi(1530)}-m_{\Sigma(1385)}.\eqno(7.5c)$$

\noindent
All of these are well satisfied: the right and left hand
sides of (7.5a) being 293 MeV and 307 MeV respectively,
of (7.5b) being 192 MeV and 212 MeV respectively, and
of (7.5c) being 147 MeV and 148 MeV respectively.
The Gell-Mann-Okubo relation

$$3m_\Lambda+m_\Sigma=2(m_N+m_\Xi), \eqno(7.6)$$

\noindent
and the equal spacing rules

$$m_\Omega-m_{\Xi(1530)}=
m_{\Xi(1530)}-m_{\Sigma(1385)}=
m_{\Sigma(1385)}-m_\Delta, \eqno(7.7)$$

\noindent
are recovered in the $SU(3)_F$ symmetric limit of the
chiral boson exchange interaction.\\

The contributions to the baryon resonances in the $N>0$
bands from the chiral boson exchange interaction (3.2)
are listed in Tables 6-10.
The lowest excited states with $N>0$ in the nucleon
and $\Delta$ spectra are the $N=2$, $L=0$
breathing modes.
The relevant integrals $P_{20}^k$
and the
oscillator parameter $\hbar\omega$ can be determined from
the $N(1440)-N$, $\Delta(1600)-N$ and  $\Lambda(1600)-N$
mass differences. As in the case of the ground state matrix
elements we take $P_{20}^{us} = P_{20}^K$
and thus there are only two independent
radial matrix elements
 for each shell. The other $\eta$ - exchange matrix
elements are determined by the expressions (7.1).
For the $P_{00}$ integrals and constituent masses
we use the second set of parameter values above.
These splittings lead to the relations

$$m_{N(1440)}-m_N$$
$$ =
{15\over 2}P_{00}^\pi-{1\over 2}P_{00}^{uu}
-{15\over 2}P_{20}^\pi+{1\over 2}P_{20}^{uu}+2\hbar\omega,\eqno (7.8a)$$

$$m_{\Delta(1600)}-m_N$$
$$=
{27\over 2}P_{00}^\pi-{3\over 2}P_{00}^{uu}
-{3\over 2}P_{20}^\pi-{1\over 2}P_{20}^{uu}+2\hbar\omega ,\eqno (7.8b)$$

$$m_{\Lambda(1600)}-m_N$$
$$=
{21\over 2}P_{00}^\pi
-{1\over 2}P_{00}^{uu}-3P_{00}^K
-{9\over 2}P_{20}^\pi+{1\over 2}P_{20}^{uu}-3P_{20}^K+2\hbar\omega
+\Delta_q ,\eqno (7.8c)$$

\noindent
which imply that
$\hbar\omega$= 156.7 MeV, $P_{20}^\pi$= 2.2 MeV, $P_{20}^K$= 0.1 MeV.
These values may be checked against the $\Sigma(1660)$, which is
the breathing
mode of the $\Sigma$:

$$m_{\Sigma(1660)}= m_N
-
{1\over 2}P_{00}^\pi
-{1\over 6}P_{00}^{uu}- {4\over 3}P_{00}^{us}-5P_{00}^K$$
$$-{1\over 2}P_{20}^\pi-{1\over 6}P_{20}^{uu}-{4\over 3}P_{20}^{us}
- 3P_{20}^K+2\hbar\omega +\Delta_q .\eqno (7.9)$$

\noindent
The result is $m_{\Sigma(1660)}= 1639$ MeV in good agreement
with the empirical value. Again if the $P_{20}$ integrals
are not determined to satisfy (7.8) exactly (in view of
the large uncertainties in the empirical values
for the masses of the resonances above considerable
freedom should be permitted in this regard) the quality
of the fit can be improved:
with  $\hbar \omega = 158.5$ MeV, $P_{20}^\pi = 3.0$ MeV,
$P_{20}^K = -2.5$ MeV we obtain the very satisfactory results
$m_{N(1440)} = 1436$ MeV, $m_{\Delta(1600)} = 1604$ MeV,
 $m_{\Lambda(1600)} = 1606$ MeV, and  $m_{\Sigma(1660)} = 1660$ MeV.
We shall use this set of parameters in the further discussion
(see also Tables 6-10).
There is no contribution to these as well nor to the
$N=0$ states from the tensor forces in first order perturbation theory.\\

	The good  quality of the prediction of this breathing
mode state suggests that the breathing mode of ${3\over 2}^+$ $\Sigma$
state should lie around  1748 MeV as seen in Table 8.
This predicted state possibly corresponds to the observed
two-star $\Sigma(1690)$ resonance the
quantum numbers of which are unknown \cite {PDG} .
In the cascade spectrum the
breathing mode states for the $\Xi$ and the $\Xi(1530)$
are predicted to be at 1798 MeV and 1886 MeV respectively.
It is difficult to make definite assignments in the cascade spectrum
since the quantum numbers of most of the identified excited states
remain unknown and several predicted states have yet to be observed
experimentally. We cannot exclude that these predicted breathing states
correspond to the
observed states $\Xi(1690)$ and $\Xi(1950)$,
which in that
case should be $\frac{1}{2}^+$ and $\frac{3}{2}^+$ states
respectively.
Because of the close similarity between the fine structure
corrections to the $N=2, L=0$ breathing mode excitations
and the corresponding states in the ground state band a mass
relation of the form (7.5b)   also applies to the breathing
resonances of the $\Sigma$'s and the $\Xi$'s:
$$m_{\Sigma(2(20)0[3]_X[3]_{FS}[3]_F[3]_S)}-
m_{\Sigma(2(20)0[3]_X[3]_{FS}[21]_F[21]_S)}$$
$$=m_{\Xi(2(20)0[3]_X[3]_{FS}[3]_F[3]_S)}
-m_{\Xi(2(20)0[3]_X[3]_{FS}[21]_F[21]_S)}.\eqno(7.10)$$

\noindent
If the assignment for $\Sigma(2(20)0[3]_X[3]_{FS}[3]_F[3]_S)$
to be the $\Sigma(1690)$ is correct,
the empirical values for the l.h.s. in (7.10) is only 30 MeV (with a large
uncertainty margin), whereas the predicted values in Table 8
gives 88 MeV. Both sets of values
 suggest that the splitting between
the breathing modes of the cascades on the r.h.s should not
exceed 100 MeV. Thus if the $\Xi(1690)$ is the breathing mode
of the $\Xi$, the $\Xi(1950)$ lies too high to be the breathing
mode of the $\Xi(1530)$ and vice versa, unless the empirical
determinations of masses of those two resonances represent
an underestimate in the case of the former, and an overestimate
in the case of the latter one.\\

The breathing mode of the $\Omega^-$ is predicted
to be at 2020 MeV. This region
of the $\Omega^-$ spectrum is predicted to contain several
states in all quark model based calculations \cite{CAI}.
Empirically the only known excited states of the $\Omega^-$
are the $\Omega^-(2250)$, $\Omega^-(2380)$ and the $\Omega^-(2470)$.
As by the present model the spectrum of the $\Omega^-$ should
be similar in structure to that of the $\Delta$, the excited
high lying $\Omega^-$ states most probably are analogs of
the corresponding highly excited $\Delta$ states above 1900 MeV,
and hence none of those represent the breathing mode of the
$\Omega^-$.\\

The two independent integrals $P_{11}^\pi$ and $P_{11}^K$
that are required for the determination
of the resonance energies in the $N=L=1$ band can be determined
from two empirical energy differences if the assumption
$P_{11}^{us}\simeq P_{11}^K$ is made as above. The large uncertainty
limits on empirical energies of the negative parity states
implies that the precision of this determination will be low.
With the spin-spin component of
the pseudoscalar exchange interaction (3.2) (for a discussion of
the tensor
and spin-orbit interactions we refer to sections 8 and 9 below) one
can try to explain qualitatively
only the centroids of the spin-flavor multiplets
in the $N=L=1$ band.
These centroid positions will be shifted by the tensor force, but
not much. There is also configuration mixing caused by the tensor
interaction and thus the assignments in Tables 3-10
imply only main components
for the baryon wave functions.
 The only multiplets in this band that are
unmixed by the tensor and spin-orbit interactions are the
$\Delta(1620)-\Delta(1700)$ doublet and the
$\Lambda(1405)-\Lambda(1520)$ flavor singlet. The excitation energies
of these states should therefore in principle be used to determine
the P-state matrix elements. A somewhat better overall
description of the baryon states in this band is however obtained
if the values of the matrix elements are chosen so as to position
the $N(1535)-N(1520)$  doublet correctly: $P_{11}^\pi=45.5$ MeV,
$P_{11}^K=30.5$ MeV.
The predicted values of other negative parity states in N=1 shell
and some of states in the $N=2,L=0$ shell which depend on
the matrix elements $P_{11}$  (e.g. $N(1710)$, $\Delta(1910)$,
$\Lambda(1810)$,...) in Tables
6 -- 9 fall  within 4\% of the corresponding empirical
values, and mostly within their uncertainty limits.
Thus the two independent matrix elements $P_{11}^\pi$
and $P_{11}^K$ suffice for a satisfactory prediction of the
energies of more than 20 confirmed
baryon states.
\\

For the final set of integrals $P_{22}$ required to complete the
table of baryon states in the $N=L=2$ band we chose
the   values $P_{22}^\pi= -35.3$ MeV and $P_{22}^K = -35.7$ MeV, as
these values lead to the correct mean energies for the
L=N=2, $[3]_{FS}[21]_F[21]_S$ nucleon and $\Lambda$ doublets
$N(1720)-N(1680)$ and $\Lambda(1890)-\Lambda(1820)$.
Again the model is supported by the good predictions
for the energies of the  confirmed $N=L=2$ band states
$\Delta(1920)$ - $\Delta(1905)$,
 $\Lambda(2110)$,$\Lambda(2020)$, $\Sigma(1915)$
and $\Sigma(2030)$.
The results in Tables 6 - 9 show
the predictions for the
energies  in the $N=L=2$ band to be
satisfactory, with the exception of the so far incompletely
determined $S={3\over 2}$ nucleon and $\Delta$ multiplets,
the only so far empirically known members of are the
$N(1990)$ and the $\Delta(1750)$ states, which are underpredicted
by 60 -- 130 MeV. As the energy of these two one star \cite {PDG}
resonances remain poorly known and both of them have widths
larger than 300 MeV, their underprediction does not at
this stage appear as a  problem for the model.\\

In Table 8 we have suggested  assignments for the
 $\Sigma$ states below 2100 MeV, for which the
spin-parity assignments are known.
Some of these assignments could have been made differently
however -- e.g. $\Sigma(1750)$,$\Sigma(2080)$ -- because of
the presence of several nearby states.
The negative
parity states $\Sigma(1940)$, $\Sigma(2000)$ and
$\Sigma(2100)$ are not included in Table 9, as they
are low lying states in the $N=3$ band. The one star
$\Sigma(1420)$, the quantum numbers of which are unknown,
is not included in the table as its
existence is uncertain \cite {PDG}.
The same applies to the two-star $\Sigma(1560)$ state,
which if confirmed probably would be a ${1\over 2}^-$
state with the flavor symmetry $[21]_F$. In this latter
case the $\Sigma(1620)$ may be the ${1\over 2}^-$
member of the multiplet $[21]_{FS}[3]_F[21]_S$.\\

For the cascade states in Table 9 we do not suggest
quantum number assignments, with exception for the
$\Xi(1820)$, as the quantum numbers of the orbital
excitations of the $\Xi$
remain unknown and several predicted states remain to
be found empirically. Given only the mass, there are several
possible assignments for each resonance.\\

The predicted excitation spectrum of the $\Omega^-$
hyperon shown in Table 10 begins around 2 GeV. The predicted
structure of the $\Omega^-$
should as mentioned above correspond to the of the $\Delta$,
and also to the $[3]_F$ parts of the spectra of the $\Sigma$
and the $\Xi$. As at least the first two of these spectra
are very satisfactorily predicted, we believe that there is
compelling reason for the existence of the predicted
$\Omega^-$ resonances around and above 2 GeV. The only
observed $\Omega^-$ resonances, the $\Omega^-(2250)$,
$\Omega^-(2380)$ and the $\Omega^-(2470)$ most probably
are all positive parity states in the $N=2$ band, and
members of the $L=0$, $S={1\over 2}$ and
$L=2$, $S={1\over 2}, {3\over 2}$ multiplets.

\vspace{1cm}

\centerline{\bf 8. The Tensor Interaction}
\vspace{1cm}

The pseudoscalar exchange mechanism that underlies the chiral
boson exchange interaction (1.1) has no spin-orbit interaction
component associated with it, and can therefore only cause a
spin-orbit splitting of the spectrum through the associated
tensor interaction. As the empirical splitting of the
baryon states in the $N=L=1$ band is small, and within the
present uncertainty limits consistent
with 0, with the exception for the anomalously large
splitting of the $\Lambda(1405)-\Lambda(1520)$ flavor singlet
spin doublet, this tensor component as well as the spin-orbit
interaction that should be associated with the scalar harmonic
confining interaction (4.1) is a priori expected to be small.
The present large uncertainties in the empirical mass values for
the baryon resonances
in the N=L=1 shell makes it difficult to determine the
strength of the tensor interaction phenomenologically
(even the sign of some of those small
spin-orbit splittings within these multiplets are not definitely
settled). Therefore we here shall evaluate the effect on
these spin-orbit splittings by the tensor interaction by
employing the Yukawa interaction model
for the pseudoscalar exchange tensor interaction.\\

The general expression for the tensor component of the
pseudoscalar octet mediated interaction is:

$$H_T = \sum_{i<j} \left\{
\sum_{a=1}^3 V_T^\pi (r_{ij}) \lambda_i^a \lambda_j^a
+\sum_{a=4}^7 V^K_T(r_{ij}) \lambda_i^a \lambda_j^a
+V^{\eta}_T(r_{ij}) \lambda_i^8 \lambda_j^8 \right\} \hat S_{ij}. \eqno(8.1)$$

\noindent
Here

$$\hat S_{ij} = 3\vec \sigma_i\cdot \hat
r_{ij}\, \vec \sigma_j\cdot \hat r_{ij}-\vec \sigma_i\cdot \vec
\sigma_j \eqno(8.2)$$

\noindent
is the tensor operator
and $V_T^{\pi}$,$V_T^{K}$ and
$V_T^{\eta}$ denote the tensor potentials that arise
from $\pi$, $K$ and
$\eta$ and  meson exchange respectively.\\

The tensor interaction (8.1) will contribute to the
energies of the
$L=1$ states for two reasons. The first is its nonvanishing and
different matrix elements for the $\frac{1}{2}^-$,$\frac{3}{2}^-$ and
$\frac{5}{2}^-$ states in the multiplets with completely symmetric
spin states ($S=\frac{3}{2}$). The second is that its nonvanishing
matrix elements between the states with different spin symmetry
but equal
flavor symmetry
causes configuration mixing
of the states with equal total angular momentum
$J$ and flavor symmetry but different
total spin $S$ in the $N=1$ band.
Thus although the diagonal matrix elements with the tensor
force for $[21]_S$ states within N=L=1 shell vanish,
some of these states will experience the tensor force
contribution through their mixing with $[3]_S$ states
that have the same J and flavor symmetry.
For example, the $N(1535)-N(1520)$ doublet the main component of which
is $\mid 1(10)[21]_X[21]_{FS}[21]_F[21]_S>$ will obtain an admixture
of the spin-quartet component $\mid 1(10)[21]_X[21]_{FS}[21]_F[3]_S>$.\\

For simplicity we adopt for this example the $SU(3)_F$
symmetric version of the spin-spin (1.1)
and tensor (8.1) components of the boson-exchange interaction.
After diagonalisation of the matrices that cause configuration mixing
of the states with $[21]_S$ and $[3]_S$ spin symmetries, the explicit
expressions for the chiral interaction contribution to the
masses of this doublet are:

$$\delta M_\chi (N(1535)) = - {9\over 2}P_{00}+{9\over 2}P_{11}
+ 4 T_{11}$$
$$- \sqrt{ (-{5\over 2}P_{00} + {1\over 2}P_{11} - 4T_{11})^2
+ 64T_{11}^2}, \eqno(8.3)$$

$$\delta M_\chi(N(1520)) = - {9\over 2}P_{00}+{9\over 2}P_{11}
- {16\over 5} T_{11}$$
$$- \sqrt{ (-{5\over 2}P_{00} + {1\over 2}P_{11} + {16\over 5}T_{11})^2
+ {32\over 5}T_{11}^2}. \eqno(8.4)$$

\noindent
Here the matrix element for the tensor interaction is defined as

$$T_{11}=<\varphi_{11m}(\vec r_{12})|V_{T}(r_{12})
|\varphi_{11m}(\vec r_{12})>.\eqno(8.5)$$

\noindent
The integrals $P_{00}$ and $P_{11}$ are taken from the section 6
to be 29.3 MeV and 45.2 MeV respectively. For this qualitative
estimate we assume the pure Yukawa radial form for the tensor interaction:

$$V_T^Y(r_{ij}) = {g^2\over {4\pi}} {{\mu^3}\over {12 m_i m_j}}
\left(1+ {3\over {\mu r_{ij}}}+ {3\over {\mu^2 r^2_{ij}}}\right)
{{exp(-\mu r_{ij})}\over {\mu r_{ij}}}. \eqno(8.6)$$

\noindent
As the $\eta$-exchange contribution
to the tensor force matrix element for $N$ and $\Delta$ states
in the N=1 band
 is suppressed
compared to the $\pi$-exchange contribution by the ratio
1:9 it suffices here to use the pure pion mass in the
Yukawa potential.\\

The coupling constant $g$ can be derived for this estimate
from the $\pi N$ coupling constant. For this we
shall use the Goldberger-Treiman relations for both the
constituent quark - pion and nucleon - pion couplings:

$$g= g^A {m_u\over f_\pi}, \eqno(8.7a)$$

$$g_{\pi N}= g^A_{\pi N} {m_N\over f_\pi}. \eqno(8.7b)$$

\noindent
Weinberg has recently shown \cite {We3} that the constituent quarks
have the bare unit axial coupling constant ($g^A =1$)
and no anomalous magnetic moment.
One thus obtains the relation

$$g= {3\over 5} {m_u\over m_N} g_{\pi N}. \eqno(8.8)$$

\noindent
The same expression can also be obtained by assuming that the
relation between the pseudovector pion-quark and pion-nucleon
coupling constants is $f={3\over 5} f_{\pi N}$.
The factor ${3\over 5}$ here and above comes from the spin-isospin
matrix element when we consider the $\pi N$ interaction as
the interaction between the pion and 3 constituent quarks.
With ${g_{\pi N}^2\over {4\pi}} = 14.2$ one has ${g^2\over {4\pi}}= 0.67$.\\

The matrix element (8.5) with the potential (8.6) is then

$$<\varphi_{11m}(\vec r_{12})|V_{T}^Y(r_{12})
|\varphi_{11m}(\vec r_{12})>$$
$$={g^2\over {4\pi}}{{\mu^3}\over {12 m_u^2}}\sqrt{2\over \pi}
\left[-\frac {1}{3b\mu} + \frac {b\mu}{3}+\frac {1}{b^3{\mu}^3}
-{1\over 3}b^2 {\mu}^2 \sqrt{{\pi \over 2}} exp( \frac {b^2 {\mu}^2} {2})
erfc( \frac {b \mu}{\sqrt{2}}) \right], \eqno(8.9)$$

\noindent
where $b$ is nucleon mean-square radius for which we take the value
$b=0.86$ fm  (see section 6).
This yields the value
$T_{11} \simeq 4.2$ MeV,
which is much smaller than the corresponding radial matrix elements
of the
spin-spin interaction.
The contribution from the tensor force will become even
smaller by the natural regularization effect of the
the finite size of the constituent quarks and the pseudoscalar
meson. Any
vector-octet-like exchange interaction component
between the constituent quarks,
would also reduce the net tensor interaction at short range as
the contributions to the tensor interaction from
pseudoscalar and vector exchange mechanisms tend to cancel,
whereas they add in the case of the spin-spin component.
These modifications of the the tensor interaction
at short range may
even lead to a sign change of the matrix element (8.9),
but in any case to a smaller absolute value than above.\\

The contribution from the tensor
forces to the baryons with spatial structure
$\mid 0 (00) 0 [3]_X>$ and $\mid 2 (20) 0 [3]_X>$
 is identically zero in first order perturbation theory.
The tensor force will however cause a
 small admixture of an $L=2$ component
in the ground state wave functions as well as in the breathing mode
states when  the calculations are performed beyond the first order
perturbation theory. Such D-wave admixtures
will bring along a small quadrupole
moments for the spin quartet states ($[3]_S$) and
are responsible for the observed $E2$ $N \to
\Delta$ transition.\\

This estimate for the matrix element $T_{11}$ implies
a small splitting in the $N(1535) - N(1520)$ doublet,
$m_{N(1535)} - m_{N(1520)} = -6.4$ MeV, and a downshift of its
centroid by 4.7 MeV.
The admixture of the $[3]_S$ state in the  $N(1535)$ and $N(1520)$
wave functions is 5.2\% and 1.9\% respectively.
At the same time the centroid of
the $N(1650) - N(1700) - N(1675)$ triplet is shifted up
by 7.6 MeV to 1640 MeV, which lies within the uncertainty
limits of the empirical value.
The same
result also applies to the $\Lambda$ spectrum and as
seen from the Table 4 the tensor force shifts the
$\Lambda(1670) - \Lambda(1690)$ doublet and the
$\Lambda(1800) - \Lambda(?) - \Lambda(1830)$ triplet in
the right directions.\\

Thus taking into account the tensor interaction component of
the pseudoscalar exchange interaction actually leads to a small
improvement of the predicted baryon spectrum.
The tensor forces do not contribute to the states
$\mid 1(10)1[21]_X[21]_{FS}[3]_F[21]_S>$. This implies that
the splitting of the $\Delta(1620) - \Delta(1700)$ doublet
should vanish in first order perturbation theory,
 a prediction which is consistent with
observation because of the large uncertainty in the corresponding
empirical mass values.\\

\vspace{1cm}

\centerline {\bf 9. The ${\bf\Lambda(1405)
- \Lambda(1520)}$ Splitting}
\vspace{1cm}

Although the pseudoscalar exchange
interaction considered above provides an explanation of
the relatively low energy of the centroid of the
$\Lambda(1405)-\Lambda(1520)$ flavor
singlet the tensor interaction that is expected
to be associated with it cannot
explain its spin-orbit splitting in first order
perturbation theory as its matrix element for that
state vanishes.
The exceptionally large spin-orbit splitting of the
$\Lambda(1405)-\Lambda(1520)$ flavor singlet suggests
a dynamical origin that is specific to that state.
That is to be expected a priori, as the $\Lambda(1405)$,
which lies slightly
below the $\overline {K} N$ threshold, may be described
as a $\overline {K} N$
bound state \cite{DAL1,LAW,SCHN,SIE,DAL2}. That implies
that it has an appreciable 3 quark + octet meson component
in addition to the basic 3 quark component and that therefore
the chiral meson field cannot be completely integrated out in
the case of $\Lambda(1405)$ as in the other baryons.\\

In this section we investigate the other possibility that
the large
spin-orbit splitting in this case might be ascribed to the
effective vector-meson-exchange
like interactions, that are naturally expected to
arise in the second iteration of the pseudoscalar exchange
interaction.\\

The fact that the lowest ${1\over 2}^-$ $\Lambda$
state lies below
the lowest  ${3\over 2}^-$ $\Lambda$ state indicates
that the spin-orbit
splitting cannot be due to the spin-orbit interaction that
is associated with the scalar confining interaction alone,
as that would lead to the opposite ordering as in the
case of the corresponding spin-orbit interaction component of
the nucleon-nucleon interaction (see also \cite{GRO}). To obtain
a spin-orbit splitting that gives a lower energy for the
${1\over 2}^-$ than for the ${3\over 2}^-$ states,
and thus the right sign for the $\Lambda(1405)-
\Lambda(1520)$ splitting
one therefore has to  invoke the spin-orbit component that is
associated by exchange of the vector  octet between
the constituent quarks.
Inclusion of that vector
meson-like spin-orbit interaction
leads to a total spin-orbit interaction of the form

$$H_{LS}=-\sum_{i<j}
\frac {1}{2}(\vec \sigma_i +\vec \sigma_j)\cdot \vec
L_{ij}$$
$$\times  \left\{V_{LS}^S(r_{ij})+
\sum_{a=1}^3 V_{LS}^{\rho}(r_{ij})
\lambda_i^a\cdot\lambda_j^a
+\sum_{a=4}^7 V_{LS}^{K^*}(r_{ij})
\lambda_i^a\cdot\lambda_j^a
+V_{LS}^{\omega}(r_{ij})\lambda_i^8\lambda_j^8\right\}
.\eqno(9.1)$$

\noindent
Here $\vec L_{ij}$ is the orbital momentum of the relative
motion of the quark pair {ij}.
The potential $V_{LS}^S$ denotes the spin-orbit interaction that
arises from the scalar confining interaction and the potentials
$V_{LS}^\rho$, $V_{LS}^{K^*}$ and $V_{LS}^{\omega}$ denote those
that arise from exchanges of systems with the quantum numbers
of the $\rho$, $K^*$ and $\omega$
mesons respectively.\\

The large splitting in the $\Lambda(1405) - \Lambda(1520)$
doublet implies that this spin-orbit force should be strong.
A vector-octet-like spin-orbit interaction that were sufficiently
strong to explain the large splitting of the flavor doublet above
would however also lead to large -- and empirically
contraindicated -- spin-orbit
splittings for the other multiplets in the $N=L=1$ band.
The question is therefore
whether or not the effect of such a large spin-orbit interaction
can be compensated
by the tensor force in the case of other multiplets.\\

The spin-orbit splittings of the multiplets in the
$N=L=1$ band of the spectrum can be expressed in terms of the
following integrals of the spin-orbit and tensor potentials
defined in (9.1) and (8.1):

$$V_{11}^k=<\varphi_{11m}|V_{LS}^k(r_{12})
|\varphi_{11m}>,\eqno(9.2a)$$

$$T_{11}^k=<\varphi_{11m}|V_{T}^k(r_{12})
|\varphi_{11m}>.\eqno(9.2b)$$

\noindent
The explicit expressions for the matrix elements of the tensor
and spin-orbit potential for the
the baryon states in the
$N=L=1$ band
that arise from the spin-spin interaction (3.3),
the spin-orbit interaction (9.1) and the tensor force (8.1)
are listed in Table 11.
The asterisk (*) on the matrix elements in the table
indicate that they are the net matrix elements for the $N,\Lambda,
\Sigma$ and $\Xi$ sectors that are defined in Table 12. \\

After diagonalization of the matrices in Table 11
that cause configuration mixing of
states in the different multiplets with $[21]_S$ and $[3]_S$ spin
symmetry
the explicit expressions for the spin-orbit splitting
of the $\frac{1}{2}^-$,$\frac{3}{2}^-$  doublet states are

$$\delta_{[111]}(\frac{1}{2}-\frac{3}{2})
= - 3 V_{11}^*, \eqno(9.3a),$$

$$\delta_{[21]}(\frac{1}{2}-\frac{3}{2})=
-\frac{3}{2}V_{11}^*+\frac{36}{5} T_{11}^*$$
$$+\frac{1}{2}
\sqrt{\left( \Delta
-\frac{3}{2}V_{11}^*-\frac{32}{5}T_{11}^*\right)^2
+40 \left( \frac{4}{5}T_{11}^*+\frac{1}{4}V_{11}^*\right)^2}$$
$$- \frac{1}{2} \sqrt{\left( \Delta
-\frac{3}{2}V_{11}^*+8T_{11}^*\right)^2
+4\left(-8T_{11}^*+\frac{1}{2}V_{11}^*\right)^2}.\eqno(9.3b)$$

\noindent
Here the flavor-symmetry is indicated by the subscripts on
the splittings and
$\Delta$ is
$$\Delta =\delta M_\chi([21]_F[3]_S)
-\delta M_\chi ([21]_F[21]_S),\eqno(9.4)$$

\noindent
where the corresponding corrections
$\delta M_{\chi}$ to the energies of those
states from the spin-spin interaction (3.2) are listed in
Tables 7-10 for the different baryon sectors.
The first splitting (9.3a) is that for the
$\Lambda(1405)-\Lambda(1520)$ doublet, which remains
unmixed. The latter (9.3b) is that for the $N(1535)-N(1520)$,
$\Lambda(1670)-\Lambda(1690)$
and corresponding $\Sigma$ and $\Xi$
doublets, which will obtain $S=\frac{3}{2}$ components
by configuration mixing.\\

The corresponding spin-orbit splittings
between the $\frac{1}{2}^-$
and $\frac{3}{2}^-$ and the $\frac{1}{2}^-$ and $\frac{5}{2}^-$
states in the triplets are

$$\delta_{[21]}(\frac{1}{2}-\frac{3}{2})=
-\frac{3}{2}V_{11}^*+\frac{36}{5}T_{11}^*$$
$$-\frac{1}{2}
\sqrt{\left(\Delta
-\frac{3}{2}V_{11}^*-\frac{32}{5}T_{11}^*\right)^2
+40\left(\frac{4}{5}T_{11}^*+\frac{1}{4}V_{11}^*\right)^2}$$
$$+ \frac{1}{2} \sqrt{\left(\Delta
-\frac{3}{2}V_{11}^*+8T_{11}^*\right)^2
+4\left(-8T_{11}^*+\frac{1}{2}V_{11}^*\right)^2},\eqno(9.5a)$$

$$\delta_{[21]}(\frac{1}{2}-\frac{5}{2})
=-\frac{1}{2}\Delta
-\frac{13}{4}V_{11}^*+\frac{12}{5}T_{11}^*$$
$$+\frac{1}{2}
\sqrt{\left(\Delta -\frac{3}{2}V_{11}^*+8T_{11}^*\right)^2
+4\left(-8T_{11}^*+\frac{1}{2}V_{11}^*\right)^2}.\eqno(9.5b)$$

\medskip

\noindent
Below for simplicity we again assume the $SU(3)_F$
limit and take $V_{LS}^\rho\simeq V_{LS}^\omega \simeq V_{LS}^{K^*}$.\\

To get the large  spin-orbit splitting 115 MeV for the
doublet $\Lambda(1405)-\Lambda(1520)$
would require that the effective matrix element
$V_{11}^*$ in the $\Lambda$ sector be as large as 38 MeV.
 At the same time the centroid of this doublet is shifted down
by $0.5 V_{11}^*$, which is favourable.
To maintain the small spin-orbit splittings of the mixed
flavor symmetry $[21]_F$  doublet and  triplet
states would then require that  it be balanced by
a correspondingly large tensor interaction matrix element
$T_{11}^*$. Sufficiently small net splittings of those states
can in principle
be obtained if $T_{11}^*$ is taken to be 13 MeV, but only
at the price of shifts of the order 27 MeV and 9 MeV down
of the centroids of the $[21]_S$ and $[3]_S$
multiplets respectively.
This downshift
can however be
compensated  by increase of the  $P_{11}$ matrix element
by a few MeV. Thus the only criterium here can be the empirical
separation between the centroids of the  $S={1\over 2}$ and $S={3\over 2}$
multiplets. With such large values for $V_{11}^*$ and $T_{11}^*$
this separation is about 120 MeV
which should be compared with the corresponding empirical separation of 148
MeV in the nucleon sector and of 135 MeV in the $\Lambda$ sector. \\

If on the other hand the strength of the spin-orbit potential
is taken to be smaller
, $V_{11}^*\simeq 25 $ MeV, a value that explains
most (75 MeV) of the $\Lambda(1405)-\Lambda(1520)$ splitting,
a much
weaker tensor interaction is required to compensate
the unfavorable splittings
in the $[21]_F$ multiplets. For example, with $\hbar \omega = 180$ MeV,
$P_{00} = 27.4$ MeV and $P_{11} = 50$ MeV and taking $T_{11}^*$
to be 7 MeV we get  the centroids of the  $N(1535) - N(1520)$
and $\Lambda(1670) - \Lambda(1690)$ doublets
at 1518 MeV and 1694 MeV respectively,
 the centroids of the $N(1650) - N(1700) - N(1675)$
and $\Lambda(1800) - \Lambda(?) - \Lambda(1830)$ triplets around
1660 MeV and 1835 MeV respectively,
and the following splittings within these multiplets:
$m_{N(1535)} - m_{N(1520)} =$ $m_{\Lambda(1670)} - m_{\Lambda(1690)} = -18$
 MeV,
$m_{N(1650)} - m_{N(1675)} =$ $m_{\Lambda(1800)} - m_{\Lambda(1830)} =
 -40$ MeV, $m_{N(1650)} - m_{N(1700)} =$ $m_{\Lambda(1800)} -
m_{\Lambda(?)} =
 44$ MeV.\\

Thus an at least qualitative explanation of the existing
spin-orbit splittings can in principle be achieved with the
assumption of a sizeable vector meson octet like interaction
between
the constituent quarks.
Attempting a quantitative explanation of the
the  spin-orbit
splittings in this way may of course require going beyond
first order
perturbation theory. \\

As noted above this qualitative explanation above for the larger part
the spin-orbit splitting
of the $\Lambda(1405) - \Lambda(1520)$ doublet is of course only
suggestive. Several other possible mechanisms
may generate large spin-orbit splittings of  the flavor-singlet
baryons and at the same time small (or vanishing) ones for baryons
with mixed or complete flavor symmetry. One such mechanism is
the short range instanton-induced
three - quark 't Hooft interaction
\cite {THO}, which involves all three flavors simultaneously
in the totally antisymmetric state and which does not
contribute to states with mixed or complete flavor symmetry.
Finally, there remains the much discussed possiblility
that the $\Lambda(1405)$ and the $\Lambda(1520)$
states contain appreciable 5 quark components, as implied e.g. by the
bound state soliton model \cite{CAL}, which automatically
leads to a large (100 -- 200 MeV) spin-orbit splitting
for that doublet \cite{BLO,RIS}.

\vspace{1cm}

{\bf 10. Exchange Current Corrections to the Magnetic Moments}
\vspace{1cm}

A flavor dependent interaction of the form (1.1) will imply the
presence of an irreducible two-body exchange current
operator, as seen e.g. directly from the continuity equation,
by which the commutator of the interaction and the single
particle charge operator equals the divergence of the exchange
current density \cite{RISK}. Because this commutator vanishes
with interparticle separation \cite{SACHS} this exchange current
is however a priori expected to be of less importance for baryons,
than for nuclei, in which the longer range of the wave functions
can lead to large matrix elements of the pion exchange current
operator.
This is one contributing reason for why the naive constituent quark
model provides such a successful description of the magnetic moments.
There has nevertheless been considerable
discussion of the  pion exchange current operator for quark pair
states in the
literature \cite{HYUGA,VENTO,DROB,BUCH}.\\

The general form of the octet vector exchange current operator
that is associated
with the complete octet mediated interaction (3.2) will have the form

$$\vec \mu^{ex}=\mu_N\{\tilde{V}_\pi(r_{ij})(\vec \tau_i\times \vec
\tau_j)_3$$

$$+\tilde{V}_K(r_{ij})(\lambda_i^4\lambda_j^5-\lambda_i^5
\lambda_j^4)\}(\vec\sigma_i \times \vec \sigma_j).\eqno(10.1)$$

\noindent
Here $\tilde{V}_\pi(r)$ and $\tilde{V}_K(r)$ are dimensionless
functions that describe $\pi$ and $K$ exchange respectively.
At long range where the interaction between quarks can be described by
a pure Yukawa potential the
function $\tilde{V}_\pi(r)$ approaches the pion exchange form

$$\tilde{V}_\pi(r_{ij})\rightarrow {g^2\over 4\pi}{1\over 3}{\mu m_N\over
m_im_j}(2\mu r_{ij}-1){e^{-\mu r_{ij}}\over \mu r_{ij}},\eqno(10.2)$$

\noindent
which includes both the pionic current and the pair current term.\\

The exchange current operator (10.1) will give rise to the following
corrections
to the magnetic moments of the ground state baryon octet:

$$\mu^{ex}(p)=-\mu^{ex}(n)=
-4<\varphi_{000}(\vec r_{12})
|\tilde{V}_\pi(r_{12})|\varphi_{000}(\vec r_{12})>
\mu_N ,\eqno(10.3a)$$

$$\mu^{ex}(\Lambda)=-\mu^{ex}(\Sigma^0)=2<\varphi_{000}(\vec r_{12})
|\tilde{V}_K(r_{12})|\varphi_{000}(\vec r_{12})>\mu_N
,\eqno(10.3b)$$

$$\mu^{ex}(\Sigma^+)=-\mu^{ex}(\Xi^0)=-4<\varphi_{000}(\vec r_{12})
|\tilde{V}_K(r_{12})|\varphi_{000}(\vec r_{12})>\mu_N
,\eqno(10.3c)$$

$$\mu^{ex}(\Sigma^-)=\mu^{ex}(\Xi^-)=0,\eqno(10.3d)$$

$$\mu^{ex}(\Sigma^0\rightarrow \Lambda)=-{4\over
\sqrt{3}}
<\varphi_{000}(\vec r_{12})
|\tilde{V}_\pi(r_{12})|\varphi_{000}(\vec r_{12})>\mu_N
$$

$$-{2\over
\sqrt{3}}<\varphi_{000}(\vec r_{12})
|\tilde{V}_K(r_{12})|\varphi_{000}(\vec r_{12})>\mu_N
.\eqno(10.3e)$$

\noindent
Here the notation for the matrix elements is the same as in eq. (5.6).
The exchange current operator (10.1) cannot contribute to the
magnetic moments of the ground state decuplet baryons, which have
completely symmetric flavor and spin states. The absence of an
exchange current correction to the magnetic moments of the $\Sigma^-$
and $\Xi^-$ is an immediate consequence of the fact that they are
formed only of $d$ and $s$ quarks, which have equal charge.\\

The impulse approximation expressions for the magnetic moments of the
ground state octet baryons and their experimental values are listed in
Table 13. If these expressions are used to
determine the mass ratios $m_N/m_u$ and
$m_N/m_s$ so as to reproduce the experimental values of
the magnetic moments of the proton and the $\Lambda$ (i.e.
$m_N/m_u=2.79,\,\,m_N/m_s=1.83$), the quark model predictions for the
$\Sigma^-$ and the cascade hyperons
as well as for those decuplet states, the
magnetic moments of which are experimentally known
($\Omega$ and $\Delta^{++}$)
will differ from the experimental
values by 15-30\% \cite{COMM}. These values for the mass ratios
moreover imply that the quark mass difference $\Delta_q$ should be 183
MeV, which is much larger than the values $\Delta_q\simeq$ 130
MeV required by the spectrum (Table 5).\\

A more natural approach is to determine the mass ratios $m_N/m_u$ and
$m_N/m_s$ to fit the experimental values of the
magnetic moments of the
$\Sigma^-$ and $\Xi^-$ octet and   the
$\Omega$ and $\Delta^{++}$ ($\mu_{\Omega}=-2.019 \pm 0.054 ~\mu_N$
\cite{Walla}, $\mu_{\Delta^{++}}= 4.52\pm 0.50 ~\mu_N$ \cite{Boss})
decuplet baryons,
which are unaffected by the exchange current operator (10.1).
While with only two independent variables it is not possible to fit
all four experimental magnetic moments exactly, the
best overall fit

$$\mu_{\Sigma^-} = -1.00 ~\mu_N,$$
$$\mu_{\Xi^-} = -0.59 ~\mu_N,$$
$$\mu_{\Omega^-} = -2.01 ~\mu_N,$$
$$\mu_{\Delta^{++}} = 5.52 ~\mu_N$$

\noindent
happens to be obtained with precisely the ratios
$m_N/m_u=2.76$ and $m_N/m_s=2.01$,
which used for constituent quark masses
to fit baryon spectrum in section 7 ($m_u = 340$ MeV and $m_s = 467$ MeV).
\\

With the given value for the strange quark mass the impulse value
for the
magnetic moment of the $\Lambda$ hyperon is
$\mu_{\Lambda}^{IA} = -0.67 ~\mu_N$. As this exceeds the experimental
value $-0.61\mu_N$ by only 10\% in magnitude the implication is
that the K-exchange radial matrix element in (10.3b) should be no
larger than
$<\varphi_{000}(\vec r_{12})
|\tilde{V}_K(r_{12})|\varphi_{000}(\vec r_{12})> = 0.03$. The meson
(kaon) - exchange contribution to the magnetic moment of $\Lambda$
thus does not exceed  10\%.\\

With the light quark mass value
above the differences between the experimental
proton and neutron magnetic moments and the corresponding
impulse approximation predictions are also
very small: $\mu_p^{exp} - \mu_p^{IA} = 0.03 ~\mu_N$ and
$\mu_n^{exp} - \mu_n^{IA} = -0.07 ~\mu_N$. This
implies that the
pion-exchange current contribution
should be very small, and that the corresponding radial integral
in (10.3a) should be
$<\varphi_{000}(\vec r_{12})
|\tilde{V}_\pi(r_{12})|\varphi_{000}(\vec r_{12})> = -(0.008 - 0.018)$.\\

With the negative value for $<\varphi_{000}(\vec r_{12})
|\tilde{V}_\pi(r_{12})|\varphi_{000}(\vec r_{12})>$ the meson-exchange
current contribution improves the theoretical value for the
$N \to \Delta$ transition,

$$\mu^{ex}(N \to \Delta)=
-4 \sqrt{2} <\varphi_{000}(\vec r_{12})
|\tilde{V}_\pi(r_{12})|\varphi_{000}(\vec r_{12})> ~\mu_N$$

$$\simeq (0.045 - 0.102) ~\mu_N,\eqno(10.4)$$

\noindent
somewhat. The sign of this MEC correction is opposite
to that  found with a pure Yukawa model for a pion-exchange interaction
in ref. \cite{DROB}. This again confirms the crucial importance
of the smearing of a $\delta$-function term in the Yukawa potential.
The value above is however not large enough to explain the whole
difference between the experimental transition magnetic moment,
$\mu^{exp}(N \to \Delta)= 3.1 - 3.2 ~\mu_N$, and the impulse (one body)
contribution,
$\mu^{IA}(N \to \Delta)= \frac {2 \sqrt{2}}{3}~ \mu_N = 2.6~\mu_N$.\\

With the values for the $\pi$- and $K$-exchange contributions
extracted above
the results in Table 13 show that the predictions for the
magnetic moments of the other octet baryon also are
improved as compared to the impulse approximation results.
The present phenomenological
analysis suggests that the meson exchange current contribution to the
octet magnetic moments does not exceed 10\%, which agrees with
the expectation above.
As the discussion here has
been purely phenomenological it has however left open the task of
constructing a model potential functions $V_\pi$ and
$V_K$ in (3.3), which have the matrix elements
$P_{nl}^k$ as required by the spectrum, and which should lead to
associated exchange current operators with radial behavior
$\tilde{V}_\pi(r)$ and
$\tilde{V}_K(r)$, with $S$-state matrix elements in the
oscillator basis of the required magnitude. The construction of the
radial part of the exchange current operator
from the interaction potential can in principle be
carried out
using the methods of ref.
\cite{RISPHY}.\\

\vspace{1cm}

\centerline{\bf 11. Discussion}
\vspace{1cm}

The agreement between the empirical baryon spectra and
those predicted above
treating the chiral boson exchange interaction (3.2)
in first order perturbation theory is quite remarkable.
While it should suffice to prove that the structure of the
interaction mediated by the pseudoscalar octet of
Goldstone bosons is essential for the understanding of
the fine structure of the spectrum it also suggests that
the baryon spectrum can be understood in the following
way.\\

If the approximate chiral symmetry of the underlying QCD were
realized in the explicit (Wigner-Weyl) mode, all hadron states
should appear with nearby parity partners. The fact that
the low lying part of both the meson and the baryon spectra
lack this feature thus implies that the approximate
chiral symmetry is spontaneously broken
and realized in the hidden (Nambu-Goldstone) mode. This
implies the generation of the dynamical mass of the
valence quarks and the presence of octet of pseudoscalar
Goldstone bosons, which are coupled directly to the constituent
quarks. In this low energy region
the gross
structure of the spectrum is caused by the confining
interaction, and (most of) the fine structure by the
interaction (1.1) (or (3.3))
that is mediated by the
octet of pseudoscalar Goldstone bosons, which are
associated with the hidden mode of chiral symmetry. \\

Without the chiral interaction, the harmonic confining interaction
would organize the baryon spectrum into equidistant shells
of alternating parity. The chiral boson exchange interaction
between the constituent quarks shifts some of the negative
parity states in the N=1 shell and positive parity states
in the N=2 shell towards each other, which leads to approximate
parity doublets. Among these are the ${1\over 2}$ near
parity doublet $\Lambda(1800)-\Lambda(1810)$, the
${3\over 2}$ parity doublet $N(1700)-N(1720)$ and the
${5\over 2}$ parity doublets $N(1675)-N(1680)$ and
$\Lambda(1830)-\Lambda(1820)$. This demonstrates the role
of the pseudoscalar interaction for partial restoration of
chiral symmetry.\\

The role of the  pseudoscalar mesons for the partial
restoration of chiral symmetry was recognized early on.
Thus the continuum states formed of the baryons and odd
numbers of these pseudoscalar mesons form the approximate
parity partners of the low lying baryon states. The
divergence of the axial current does not vanish as in the
explicit mode of chiral symmetry but it is proportional
to the pseudoscalar meson field (PCAC) and does vanish
in the  limit $m_\pi \to 0$. The smallness of the
breaking of the underlying chiral symmetry is revealed
by the remarkable accuracy of the Goldberger-Treiman
relation for the pion-nucleon coupling. The present model
achieves the partial restoration of chiral symmetry at
a more microscopic level, which leads to the explanation
of the appearance of the near parity doublets in the
spectrum.\\

The
(still poorly mapped) high energy
part of the baryon spectrum on the other hand, which is formed of
a gradually increasing number of near
degenerate parity doublets (or more generally multiplets),
should reveal the
explicit Wigner-Weyl mode of chiral symmetry, which is due to the
indistinguishability between left- and right-handed massless quarks
in QCD. The remaining small splitting of the degeneracy between the
parity partners is then due to the small mass
of the current quarks and the gradually vanishing hidden mode of
chiral symmetry. \\

The baryon spectrum suggests that the
phase transition between
the Nambu-Goldstone and Wigner-Weyl mode of chiral symmetry is
gradual, as there already in the hidden mode appears a partial
restoration of chiral symmetry and as
the mass difference between the nearest neighbors with
opposite parity falls to zero only gradually with increasing
resonance energy. The clearest signal for this is that while
the
splitting within the $\Lambda(1600)-\Lambda(1670)$ parity
doublet is still 70 MeV, the splittings within the $J^P=
\frac{1}{2}^\pm$ and $J^P=\frac{5}{2}^\pm$ $\Lambda$-resonance
parity doublets
around 1800 MeV are only 10 MeV.
This is an indication of the amorfic (disordered) structure of
the QCD vacuum and its quark condensate. The implication
would then be that
there is a
gradual chiral restoration phase transition.
The disordered quark condensate structure of the QCD vacuum
appears in the instanton liquid  model of the QCD vacuum
\cite{SHU,DIA}. Because of the gradual character of this
phase transition no definite transition energy can be defined.
If
the onset
of the parity doubling in the resonance spectrum is taken to be
at about
500 MeV above the ground state, as suggested by the mass difference
between the $\Lambda$ and the lowest parity doublet formed by the
$\Lambda(1600)$ and the $\Lambda(1670)$, or by the mass
difference between the nucleon and the $N(1440)-N(1535)$
pair,
the approximate transition energy would
then by of the order 500 MeV.
The absence of
structure in the baryon spectrum above 2 GeV excitation energy
suggests that
the Nambu-Goldstone mode has totally
disappeared in that
energy range.\\

The present results indicate that the role of the one gluon
exchange
interaction, which should be important in the Wigner-Weyl mode
and for current quarks,
for the ordering of the baryon spectrum is small. If
it is included in the model as a phenomenological term the value of
the effective coupling strength $\alpha_S$ should be much smaller than
the values $\sim 1$ that have been typically employed
\cite{IGK1,IGK2,CAI}.\\

Quark-quark interactions
that involves the flavor degrees of freedom have been
found to arise in the instanton induced interaction between quarks \cite{THO}.
This interaction taken between
constituent quarks
has recently been applied directly to baryon structure
\cite {DOR,SHU2,TAK,TAK2}.
It differs
in a crucial aspect from the
pseudoscalar octet mediated interaction (1.1) in that it
vanishes in flavor symmetric pair states.
As a consequence
it fails to account for the fine structure in the
$\Delta$-spectrum, as exemplified e.g. in the prediction of
the wrong ordering of the $\Delta(1600)$ and the
negative parity pair $\Delta(1620)-\Delta(1700)$ \cite {DOR}.
\\

It proves instructive to consider the symmetry structure of the
harmonic confining + chiral octet mediated
interaction (1.1) model presented here in view of the
highly satisfactory
predictions obtained for the baryon spectra.
The symmetry group for the orbital part of a
harmonically bound system of A quarks is $U(3(A-1))$,
which in the present case reduces to $U(6)$. In the absence
of the fine-structure interaction (1.1), and with equal
u,d and s-
quark masses, the baryon states would form unsplit
multiplets of the full symmetry group
$SU(6)_{FS}\times U(6)_{conf}$. The $SU(3)_F$ symmetrical
version of the chiral interaction (1.1)
reduces this
degeneracy within the multiplets to those that corresponding
to $SU(3)_F\times SU(2)_S\times U(6)_{conf}$
and is in fact strong enough
to shift some of the N=2 states below the N=1 states
and to mix positions of different multiplets. Thus the N=2 resonance
$N(1440)$ is shifted down below the N=1 resonance $N(1535)$ etc.
As noted above when this shifting moves states from adjacent
N-levels close to each other near degenerate parity doublets
appear. The model thus suggests an explicit explanation of
the observed near parity doubling of the spectrum
already in the Nambu-Goldstone mode.
Within the constituent quark model the most natural suggestion
for the appearance of the near parity doublets is that the
Hamiltonian that is formed of the confining harmonic
interaction and the chiral field interaction (1.1)
contains an additional symmetry of higher rank than
$SU(3)_F\times SU(2)_S\times U(6)_{conf}$,
which combines the spatial and flavor-spin degrees of freedom.
This conjecture is supported by the relative insensitivity
of the predicted spectra to the parameter values used.
The most natural suggestion is that this "unification" is
related to the
$SU(3)_L^F\times SU(3)_R^F\times U(1)^F$ symmetry of the
underlying QCD in the Wigner-Weyl mode.\\

The mass splittings between the different members of the
same $SU(3)_F\times SU(2)_S\times U(6)_{conf}$ multiplet
arise due to both the constituent quark mass difference in
(5.4) and the different strength of the meson-exchange
interaction $V_\pi \not= V_K \not= V_\eta$ beyond the $SU(3)_F$
limit. Thus even those states in the $\Lambda$ and $\Sigma$
spectrum which have identical quark content and equal spatial,
flavor, spin and flavor-spin symmetries, get different
contributions from the interaction (3.2)-(3.3) and consequently
different masses.\\

There is no fundamental reason for why the effective confining
interaction between the constituent quarks should have to be
harmonic. The low-lying part of the baryon spectrum is  not
very sensitive to the form of the confining interaction, but
the very satisfactory numerical predictions obtained here for
the baryon spectra up to about 1 GeV excitation energy suggest
that any anharmonic corrections should be small. Quantitative
study of the detailed form of the confining interaction
would require a simultaneous specification of the detailed
short range part of the chiral interaction (1.1), and would
presumably also need increased accuracy for the empirical
resonance energies. If the harmonic confining interaction is
replaced by a nonharmonic form, the $U(6)$ spatial symmetry of the
confining form is reduced to $O(3)$.\\

The low lying part of the baryon spectrum depends to a much
higher degree on the chiral boson exchange interaction than
on the confining interaction. This can be illustrated by the
fact that only about a quarter of the mass difference between the
nucleon and the lowest ${1\over 2}^-$ state $N(1535)$ is due to
the confining central interaction, whereas the remaining
3 quarters are due to the spin-spin interaction (1.1). This
relative "weakness" of the confining interaction is the
reason
for why the oscillator parameter in the present model
is much smaller ($\simeq$ 160 MeV) than in the models that
are based on perturbative gluon exchange between the quarks.
As in the latter the color magnetic interaction (2.2) contributes
very little to the $N(1535)-N$ mass splitting the oscillator
parameter in such models is much larger ($\simeq$ 500--600 MeV).
This difference in the oscillator parameter value is the
reason for why the present chiral interaction based model
leads to the correct nucleon radius (0.86 fm), whereas the
gluon exchange based model leads to understimates ($\simeq$ 0.5 fm).
\\

Be it as it may, the present organization of fine
structure of the baryon
spectrum based on the quark-quark interaction that is
mediated by the octet of pseudoscalar mesons, which
represent the Goldstone bosons associated with the
hidden mode of the approximate chiral symmetry of QCD
is both simple and phenomenologically successful.
The predicted energies of the states in
the nucleon and strange hyperon spectra agree with the
empirical values, where known, to within a few percent.
The accuracy of the predictions can readily be improved
both by readjustment of the required matrix elements
of the fine structure interaction and by carrying the
calculation to second order.
\\

\vspace{1cm}

\centerline{\bf 12. Outlook}

\vspace{1cm}

The very satisfactory predictions obtained here for the baryon
spectrum suggest a solution to the long standing problem of finding a
quark model basis for the phenomenologically successful meson exchange
description of the nucleon-nucleon interaction.
This problem can be approached by describing the
nucleon-nucleon system as a six-quark system with the
quark-cluster ansatz for the six-quark wave function
(resonating group method (RGM) or related generator coordinate method)
\cite {OKAY},
in which the effective interaction is formed of direct
interquark interactions and quark interchanges between clusters.
Previous work based on this approach has since the early
work of Oka and Yazaki \cite {OKAY} attempted to describe the
effective repulsive short-range part of the NN system as a combination
of the
color-magnetic interaction (2.2) and the  quark
interchanges. From the present perspective it
is interesting to note that with this approach
a quantitatively satisfactory description of the nucleon-nucleon
interaction and nucleon-nucleon scattering observables requires the
presence of a pion or chiral field interaction between the quarks in
addition to the confining and possible gluon exchange terms
\cite{Ob,Fern,ROBD,Val,Zh}. The
present results indicate that the effective nucleon-nucleon
interaction can be described in terms of the chiral meson field
mediated interaction between the constituent quarks and the short range
 quark interchanges  alone, without any
need for a perturbative gluon
exchange component.
\\

In this context it is worth emphasizing that if the confining
quark-quark interaction is harmonic, it does not contribute to the
effective interaction between the three quark clusters that form the
nucleons. This is because no color Van-der-Waals forces appear in
in the one-channel RGM approach and because
the nonlocal RGM kernel that obtains
with a harmonic interquark potential is proportional to the normalizing
kernel, and therefore cancels out in the  RGM wave
equation \cite{OKAY}. It then follows that in
this approximation  the meson
exchange interaction between the quarks gives rise to a pure meson
exchange interaction between the nucleons with the
addition of  quark interchanges  at short range.
The latter appear as a consequence
of the Pauli principle at the quark level and are essential for the
short range repulsion in the two-nucleon system.
Experiments for the testing of the
quark-interchange contributions directly have been proposed in refs.
\cite{GNOS,GNO,GF,GLOZMAN}.\\

A natural final question that arises is that of the effective interaction
between quarks and antiquarks. The general relation between
meson-exchange models for the quark-quark and quark-antiquark
interactions is that their components that describe exchange of
systems with even $G$-parity have the same and those with odd
$G$-parity have the opposite sign. The discussion of the spin-orbit
interaction in section 9 above suggests that the effective meson
exchange interaction (1.1) could be formed not only of pseudoscalar
exchange but also of a vector-meson-like exchange component with the same
sign. In the $q\bar q$ channel the pseudoscalar exchange term would
have opposite sign because of the odd G-parity of the pseudoscalar
octet, and therefore the corresponding spin-spin
interaction should be much weaker because of the partial cancellation
between its pseudoscalar and vector-meson-like exchange components. On the
other hand the flavor-dependent tensor interaction (8.1), which is
weak because of the partially cancelling pseudoscalar and vector
exchange components should be much stronger in the $q\bar q$-channel,
in which the two components have the same sign. As a consequence of
this the effective flavor dependent meson exchange interaction between
quarks and antiquarks should be very different from that between
quarks. Hence no immediate conclusion concerning the spectrum of the
vector - and heavier meson can be drawn from the present results. No
simple two-particle interaction model should in any case
be expected to apply to
the lowest pseudoscalar mesons, which are approximate Goldstone bosons
and thus collective $q\bar q$ excitations but
not simple two-body $q\bar q$-systems.\\


\
\newpage
\centerline{\bf Table 1}

The scalar factors of the Clebsch-Gordan coefficients for the group
$SU(6)_{FS}$
in the reduction $SU(6)_{FS}\supset SU(3)_F\times SU(2)_S$
defined in eq. (4.9a).

\begin{center}
\begin{tabular}{|c|c|c|c|c|c|} \hline
\multicolumn{2}{|c|}{}&\multicolumn{2}{c|}{}&\multicolumn{2}{c|}{}\\
\multicolumn{2}{|c|}{}&\multicolumn{2}{c|} {$[f_{12}]_{FS}=[2]$}
&\multicolumn{2}{c|} {$[f_{12}]_{FS}=[11]$}\\
\multicolumn{2}{|c|}{}&\multicolumn{2}{c|}{}&\multicolumn{2}{c|}{}
\\ \hline
&&&&&\\
$[f]_{FS}$&$[f]_F S$&$[f_{12}]_F S_{12}=[2]1$&
$[f_{12}]_F S_{12}=[11]0$&$[f_{12}]_F S_{12}=[2]0$&
$[f_{12}]_F S_{12}=[11]1$\\
&&&&&\\ \hline
&&&&&\\
&$[3]\frac{3}{2}$&$1$&&&\\
$[3]$ &&&&&\\
&$[21]\frac{1}{2}$&$\sqrt{\frac{1}{2}}$&$\sqrt{\frac{1}{2}}$&&\\
&&&&&\\ \hline
&&&&&\\
&$[3]\frac{1}{2}$&$1$&&$-1$& \\
&&&&&\\
&$[21]\frac{1}{2}$&$\sqrt{\frac{1}{2}}$&$-\sqrt{\frac{1}{2}}$&
$\sqrt{\frac{1}{2}}$&$\sqrt{\frac{1}{2}}$\\
$[21]$ &&&&&\\
&$[21]\frac{3}{2}$&$1$&&&$-1$\\
&&&&&\\
&$[111]\frac{1}{2}$&&$1$&&$1$\\
&&&&&\\ \hline
&&&&&\\
&$[21]\frac{1}{2}$&&&$\sqrt{\frac{1}{2}}$&$-\sqrt{\frac{1}{2}}$\\
$[111]$&&&&&\\
&$[111]\frac{3}{2}$&&&&$1$\\
&&&&&\\ \hline
\end{tabular}
\end{center}

\newpage
\centerline{\bf Table 2a}

Isoscalar factors of the Clebsch-Gordan coefficients for the group $SU(3)_F$
in the canonical reduction
defined in eq. (4.9b).

\begin{center}
\begin{tabular}{|c|c|c|c|c|c|} \hline
\multicolumn{2}{|c|}{}&\multicolumn{4}{c|}{}\\
\multicolumn{2}{|c|}{$[f_{12}]_F = [11]$}
&\multicolumn{4}{c|}{$Y_{12}~ T_{12}~~~ Y_3 ~ T_3$}\\
\multicolumn{2}{|c|}{}&\multicolumn{4}{c|}{}\\
\hline
&&&&&\\
$[f]_F$&$YT$&$-\frac{1}{3}\frac{1}{2}~~-\frac{2}{3}0$&
$\frac{2}{3}0~-\frac{2}{3}0$&$-\frac{1}{3}\frac{1}{2}~\frac{1}{3}
\frac{1}{2}$&$\frac{2}{3}0~\frac{1}{3}\frac{1}{2}$\\
&&&&&\\ \hline
&&&&&\\
$[21]$&$-1\frac{1}{2}$&$1$&&&\\
&&&&&\\
$[21]$&$00$&&$\sqrt{\frac{2}{3}}$&$-\sqrt{\frac{1}{3}}$&\\
&&&&&\\
$[111]$&$00$&&$\sqrt{\frac{1}{3}}$&$\sqrt{\frac{2}{3}}$&\\
&&&&&\\
$[21]$&$01$&&&$1$&\\
&&&&&\\
$[21]$&$1\frac{1}{2}$&&&&$1$\\
&&&&&\\ \hline
\end{tabular}
\end{center}

\newpage
\centerline{\bf Table 2b}

Isoscalar factors of the Clebsch-Gordan coefficients for the group $SU(3)_F$
in the canonical reduction
defined in eq. (4.9b).

\begin{center}
\begin{tabular}{|c|c|c|c|c|c|c|c|} \hline
\multicolumn{2}{|c|}{}&\multicolumn{6}{c|}{}\\
\multicolumn{2}{|c|}{$[f_{12}]_F = [2]$}
 &\multicolumn{6}{c|}{$Y_{12}~ T_{12}~~~~ Y_3~T_3$}\\
\multicolumn{2}{|c|}{}&\multicolumn{6}{c|}{}\\
\hline
&&&&&&&\\
$[f]_F$&$YT$&$-\frac{2}{3}0~\frac{1}{3}\frac{1}{2}$&
$-\frac{1}{3}\frac{1}{2}~-\frac{2}{3}0$&
$\frac{2}{3}1~-\frac{2}{3}0$&
$-\frac{1}{3}\frac{1}{2}~\frac{1}{3}\frac{1}{2}$&
$\frac{2}{3}1~\frac{1}{3}\frac{1}{2}$&
$-\frac{4}{3}0~-\frac{2}{3}0$\\
&&&&&&&\\ \hline
&&&&&&&\\
$[3]$&$-20$&&&&&&$1$\\
&&&&&&&\\
$[3]$&$-1\frac{1}{2}$&$\sqrt{\frac{1}{3}}$&
$\sqrt{\frac{2}{3}}$&&&&\\
&&&&&&&\\
$[21]$&$-1\frac{1}{2}$&$\sqrt{\frac{2}{3}}$&
$-\sqrt{\frac{1}{3}}$&&&&\\
&&&&&&&\\
$[21]$&$00$&&&&$1$&&\\
&&&&&&&\\
$[3]$&$01$&&&$\sqrt{\frac{1}{3}}$&
$\sqrt{\frac{2}{3}}$&&\\
&&&&&&&\\
$[21]$&$01$&&&$-\sqrt{\frac{2}{3}}$&
$\sqrt{\frac{1}{3}}$&&\\
&&&&&&&\\
$[21]$&$1\frac{1}{2}$&&&&&$1$&\\
&&&&&&&\\
$[3]$&$1\frac{3}{2}$&&&&&$1$&\\
&&&&&&&\\ \hline
\end{tabular}
\end{center}
\newpage

\centerline{\bf Table 3}

The structure of the nucleon
and $\Delta$ resonance states up to $N=2$ as predicted with
the $SU(3)_F$ invariant version of the chiral boson
interaction. The 11 predicted unobserved or nonconfirmed
states are indicated by question marks.
The predicted energy values (in MeV) are given in the brackets
under the empirical ones.

\begin{center}
\begin{tabular}{|llll|} \hline
$N(\lambda\mu)L[f]_X[f]_{FS}[f]_F[f]_S$
& LS multiplet & average &$\delta M_\chi$\\
&&energy&\\ \hline
$0(00)0[3]_X[3]_{FS}[21]_F[21]_S$ & ${1\over 2}^+, N$ &
939&$-14 P_{00}$\\
&&&\\
$0(00)0[3]_X[3]_{FS}[3]_F[3]_S$ & ${3\over 2}^+, \Delta$ &
1232&$-4 P_{00}$\\
&&(input)&\\
$2(20)0[3]_X[3]_{FS}[21]_F[21]_S$ & ${1\over 2}^+, N(1440)$ &
1440&$-7 P_{00}-7P_{20}$\\
&&(input)&\\
$1(10)1[21]_X[21]_{FS}[21]_F[21]_S$ & ${1\over 2}^-, N(1535);
{3\over 2}^-, N(1520)$ &
1527&$-7 P_{00}+ 5P_{11}$\\
&&(input)&\\
$2(20)0[3]_X[3]_{FS}[3]_F[3]_S$ & ${3\over 2}^+, \Delta(1600)$ &
1600&$-2 P_{00}-2P_{20}$\\
&&(input)&\\
$1(10)1[21]_X[21]_{FS}[3]_F[21]_S$ & ${1\over 2}^-, \Delta(1620);
{3\over 2}^-,\Delta(1700)$ &
1660&$-2P_{00}+6P_{11}$\\
&&(1719)&\\
$1(10)1[21]_X[21]_{FS}[21]_F[3]_S$ & ${1\over 2}^-, N(1650);
{3\over 2}^-,N(1700)$ &
1675&$-2 P_{00}+4P_{11}$\\
&${5\over 2}^-,N(1675)$&(1629)&\\
&&&\\
$2(20)2[3]_X[3]_{FS}[3]_F[3]_S$&${1\over 2}^+,\Delta(1750?);
{3\over 2}^+,\Delta(?)$&1750?&$-2P_{00}-2P_{22}$\\
&${5\over 2}^+,\Delta(?);{7\over 2}^+,\Delta(?)$&(1675)&\\
&&&\\
$2(20)2[3]_X[3]_{FS}[21]_F[21]_S$&${3\over 2}^+,N(1720);
{5\over 2}^+,N(1680)$&1700&$-7P_{00}-7P_{22}$\\
&&(input)&\\
$2(20)0[21]_X[21]_{FS}[21]_F[21]_S$ & ${1\over 2}^+, N(1710)$
&1710&$-{7\over 2}P_{00}-{7\over 2}P_{20}+5P_{11}$\\
&&(1778)&\\
$2(20)0[21]_X[21]_{FS}[21]_F[3]_S$ & ${3\over 2}^+, N(?)
$ &?&$-P_{00}-P_{20}+4P_{11}$\\
&&(1813)&\\
$2(20)2[21]_X[21]_{FS}[21]_F[21]_S$ & ${3\over 2}^+, N(1900?);
{5\over 2}^+,N(2000?);$ &1950?
&$-{7\over 2}P_{00}-{7\over 2}P_{22}+5P_{11}$\\
&&(1909)&\\
$2(20)2[21]_X[21]_{FS}[21]_F[3]_S$ & ${1\over 2}^+, N(?);
{3\over 2}^+,N(?)$&1990?
&$-P_{00}-P_{22}+4P_{11}$\\
&${5\over 2}^+,N(?);{7\over 2}^+,N(1990?)$&(1850)&\\
&&&\\
$2(20)0[21]_X[21]_{FS}[3]_F[21]_S$ & ${1\over 2}^+, \Delta(1910)
$ &1910&$-P_{00}-P_{20}+6P_{11}$\\
&&(1903)&\\
$2(20)2[21]_X[21]_{FS}[3]_F[21]_S$ & ${3\over 2}^+, \Delta(1920);
{5\over 2}^+,\Delta(1905)$ &1912
&$-P_{00}-P_{22}+6P_{11}$\\
&&(1940)&\\ \hline
\end{tabular}
\end{center}

\newpage
\centerline{\bf Table 4}
\vspace{0.5cm}

The structure of the $\Lambda$-hyperon
states up to $N=2$ predicted with the $SU(3)_F$ invariant
version of the chiral boson exchange interaction. The
10 predicted unobserved or nonconfirmed states are indicated
by question marks. The predicted energies (in MeV)
are given in the brackets under the empirical values.
\begin{center}
\begin{tabular}{|llll|} \hline
$N(\lambda\mu)L[f]_X[f]_{FS}[f]_F[f]_S$
& LS multiplet & average &$\delta M_\chi$\\
&&energy&\\ \hline
$0(00)0[3]_X[3]_{FS}[21]_F[21]_S$ & ${1\over 2}^+, \Lambda$ &
1115&$-14 P_{00}$\\
&&&\\
$1(10)1[21]_X[21]_{FS}[111]_F[21]_S$ & ${1\over 2}^-, \Lambda(1405);
{3\over 2}^-,\Lambda(1520)$ &
1462&$-12 P_{00}+4P_{11}$\\
&&(1512)&\\
$2(20)0[3]_X[3]_{FS}[21]_F[21]_S$ & ${1\over 2}^+, \Lambda(1600)$ &
1600&$-7 P_{00}-7P_{20}$\\
&&(1616)&\\
$1(10)1[21]_X[21]_{FS}[21]_F[21]_S$ & ${1\over 2}^-, \Lambda(1670);
{3\over 2}^-, \Lambda(1690)$ &
1680&$-7 P_{00}+5 P_{11}$\\
&&(1703)&\\
$1(10)1[21]_X[21]_{FS}[21]_F[3]_S$ & ${1\over 2}^-, \Lambda(1800);
{3\over 2}^-,\Lambda(?);$ &
1815&$-2 P_{00}+4P_{11}$\\
&${5\over 2}^-,\Lambda(1830)$&(1805)&\\

&&&\\
$2(20)0[21]_X[21]_{FS}[111]_F[21]_S$ & ${1\over 2}^+, \Lambda(1810)
$&1810&$-6P_{00}-6P_{20}+4P_{11}$\\
&&(1829)&\\
$2(20)2[3]_X[3]_{FS}[21]_F[21]_S$ & ${3\over 2}^+, \Lambda(1890);
{5\over 2}^+,\Lambda(1820)$ &
1855&$-7 P_{00}-7P_{22}$\\
&&(1878)&\\
$2(20)0[21]_X[21]_{FS}[21]_F[21]_S$&${1\over 2}^+,\Lambda(?)$&
?&$-{7\over 2}P_{00}-{7\over 2}P_{20}+5P_{11}$\\
&&(1954)&\\
$2(20)0[21]_X[21]_{FS}[21]_F[3]_S$ & ${3\over 2}^+, \Lambda(?)$&
?&$-P_{00}-P_{20}+4P_{11}$\\
&&(1989)&\\
$2(20)2[21]_X[21]_{FS}[21]_F[3]_S$ & ${1\over 2}^+, \Lambda(?);
{3\over 2}^+,\Lambda (?);$&2020?&
$-P_{00}-P_{22}+4P_{11}$\\
&${5\over 2}^+\Lambda(?);{7\over 2}^+,\Lambda(2020?)$&(2026)&\\
&&&\\
$2(20)2[21]_X[21]_{FS}[111]_F[21]_S$ & ${3\over 2}^+, \Lambda(?);
{5\over 2}^+,\Lambda(?)$&
?&$-6P_{00}-6P_{22}+4P_{11}$\\
&&(2053)&\\
$2(20)2[21]_X[21]_{FS}[21]_F[21]_S$ & ${3\over 2}^+,\Lambda(?);
{5\over 2}^+,\Lambda(2110)$ &2110?
&$-{7\over 2}P_{00}-{7\over 2}P_{22}+5P_{11}$\\
&&(2085)&\\ \hline
\end{tabular}
\end{center}

\newpage
\centerline{\bf Table 5}
\vspace{0.5cm}

The contributions to the masses of the baryon
states in the $N=0$  band given by the $SU(3)_F$ symmetry
breaking chiral interaction (3.2). The mass difference
between the s and u,d quarks is denoted $\Delta_q$.
The superscripts uu,us and ss on the $\eta$-exchange
matrix elements indicate that it applies to pair states
of two light, one light and on strange and two strange
quarks respectively. The
predicted mass values (in MeV) for the parameter set (I)
$\Delta_q = 121$ MeV,
$m_u = 340$MeV, $P_{00}^\pi = 28.9$ MeV, $P_{00}^K =  19.6$ MeV
and that for the set (II)
$\Delta_q = 127$ MeV,
$m_u = 340$MeV, $P_{00}^\pi = 29.05$ MeV, $P_{00}^K =  20.1$ MeV
are given in the corresponding columns.
\begin{center}
\begin{tabular}{|lllll|} \hline
$N(\lambda\mu)L[f]_X[f]_{FS}[f]_F[f]_S$
& State & Predicted & Predicted &$\delta M_\chi$\\
&(mass)& mass I& mass II&\\ \hline
&&&\\
$0(00)0[3]_X[3]_{FS}[21]_F[21]_S$ & $~N$ &
input&input&$-15 P_{00}^\pi+P_{00}^{uu}$\\
&(939)&&&\\
&&&&\\
$0(00)0[3]_X[3]_{FS}[3]_F[3]_S$ &$~~\Delta$ &1232 & 1232 &$-3 P_{00}^\pi
-P_{00}^{uu}$\\
&(1232)&&&\\
&&&&\\
$0(00)0[3]_X[3]_{FS}[21]_F[21]_S$&$~~\Lambda$&1116& 1120 &$-9P_{00}^\pi
-6 P_{00}^K + P_{00}^{uu}+\Delta_q$\\
&(1116)&&&\\
&&&&\\
$0(00)0[3]_X[3]_{FS}[21]_F[21]_S$&$~~\Sigma$ &1181&1181& $-P_{00}^\pi
-10 P_{00}^K$\\
&(1193)&&&$-{1\over 3} P_{00}^{uu}-{8\over 3}P_{00}^{us}+\Delta_q$\\
&&&&\\
$0(00)0[3]_X[3]_{FS}[3]_F[3]_S$ &$\Sigma(1385)$&1377&1382&
$-P_{00}^\pi-4 P_{00}^K$\\
&(1385)&&&$-{1\over 3}P_{00}^{uu}+{4\over 3}P_{00}^{us}+\Delta_q$\\
&&&&\\
$0(00)0[3]_X[3]_{FS}[21]_F[21]_S$&$~~\Xi$&1320&1327&$-10P_{00}^K$\\
&(1318)&&&$-{8\over 3}P_{00}^{us}-{4\over3}P_{00}^{ss}+2\Delta_q$\\
&&&&\\
$0(00)0[3]_X[3]_{FS}[3]_F[3]_S$&$\Xi(1530)$&1516&1528&$-4P_{00}^K
$\\
&(1530)&&&$+{4\over 3}P_{00}^{us}-{4\over 3}P_{00}^{ss}+2\Delta_q$\\
&&&&\\
$0(00)0[3]_X[3]_{FS}[3]_F[3]_S$&$~~\Omega^-$&1651&1670&$-4P_{00}^{ss}
+3\Delta_q$\\
&(1672)&&\\ \hline
\end{tabular}
\end{center}
\newpage
\centerline{\bf Table 6}

The structure of the nucleon
and $\Delta$ resonance states in the $N=1,2$ bands as predicted with
the $SU(3)_F$ breaking version of the chiral boson
interaction (2.3). The $\eta$-exchange matrix elements are
have the superscript uu to indicate that they apply to pair
states of light constituent quarks.
The predicted energy values (in MeV) are given in the brackets
under the empirical ones. The parameter Set is the following:
set (II) from the Table 5 plus
 $P_{11}^\pi = 45.5$ MeV, $P_{11}^K =  30.5$ MeV,
 $P_{20}^\pi = 3.0$ MeV, $P_{20}^K =  -2.5$ MeV,
 $P_{22}^\pi = -35.3$ MeV, $P_{22}^K =  -35.7$ MeV
\begin{center}
\begin{tabular}{|llll|} \hline
$N(\lambda\mu)L[f]_{FS}[f]_F[f]_S$
& LS multiplet & average &$\delta M_\chi$\\
&&energy&\\ \hline
$2(20)0[3]_{FS}[21]_F[21]_S$ & ${1\over 2}^+, N(1440)$ &
1440&$-{15\over 2} P_{00}^\pi+{1\over 2}P_{00}^{uu}$\\
&&(1436)&$-{15\over 2}P_{20}^\pi+{1\over 2}P_{20}^{uu}$\\ \hline

$1(10)1[21]_{FS}[21]_F[21]_S$ & ${1\over 2}^-, N(1535);
{3\over 2}^-, N(1520)$ &
1527&$-{15\over 2}P_{00}^\pi+{1\over 2}P_{00}^{uu}$\\
&&(1527)&$+{9\over 2}P_{11}^\pi+{1\over 2}P_{11}^{uu}$\\ \hline

$2(20)0[3]_{FS}[3]_F[3]_S$ & ${3\over 2}^+, \Delta(1600)$ &
1600&$-{3\over 2} P_{00}^\pi-{1\over 2}P_{00}^{uu}$\\
&&(1604)&$-{3\over 2}P_{20}^\pi-{1\over 2}P_{20}^{uu}$\\ \hline

$1(10)1[21]_{FS}[3]_F[21]_S$ & ${1\over 2}^-, \Delta(1620);
{3\over 2}^-,\Delta(1700)$ &
1660&$-{3\over 2}P_{00}^\pi-{1\over 2}P_{00}^{uu}$\\
&&(1716)&$+{9\over 2}P_{11}^\pi+{3\over 2}P_{11}^{uu}$\\ \hline

$1(10)1[21]_{FS}[21]_F[3]_S$ & ${1\over 2}^-, N(1650);
{3\over 2}^-,N(1700)$ &
1675&$-{3\over 2}P_{00}^\pi-{1\over 2}P_{00}^{uu}$\\
&${5\over 2}^-,N(1675)$&(1632)&
$+{9\over 2}P_{11}^\pi-{1\over 2}P_{11}^{uu}$\\ \hline

$2(20)2[3]_{FS}[3]_F[3]_S$&${1\over 2}^+,\Delta(1750?);
{3\over 2}^+,\Delta(?)$&1750?&$-{3\over 2} P_{00}^\pi
-{1\over 2}P_{00}^{uu}$\\
&${5\over 2}^+,\Delta(?);{7\over 2}^+,\Delta(?)$&(1684)&
$-{3\over 2}P_{22}^\pi-{1\over 2}P_{22}^{uu}$\\ \hline

$2(20)2[3]_{FS}[21]_F[21]_S$&${3\over 2}^+,N(1720);
{5\over 2}^+,N(1680)$&1700&$-{15\over 2} P_{00}^\pi+
{1\over 2}P_{00}^{uu}$\\
&&(1700)&$-{15\over 2}P_{22}^\pi+{1\over 2}P_{22}^{uu}$\\ \hline

$2(20)0[21]_{FS}[21]_F[21]_S$ & ${1\over 2}^+, N(1710)$
&1710&$-{15\over 4}P_{00}^\pi+{1\over 4}P_{00}^{uu}$\\
&&(1776)&$-{15\over 4}P_{20}^\pi+{1\over 4}P_{20}^{uu}$\\
&&&$+{9\over 2}P_{11}^\pi+{1\over 2}P_{11}^{uu}$\\ \hline

$2(20)0[21]_{FS}[21]_F[3]_S$ & ${3\over 2}^+, N(?)
$ &?&$-{3\over 4}P_{00}^\pi-{1\over 4}P_{00}^{uu}$\\
&&(1818)&$-{3\over 4}P_{20}^\pi-{1\over 4}P_{20}^{uu}$\\
&&&$+{9\over 2}P_{11}^\pi-{1\over 2}P_{11}^{uu}$\\ \hline
$2(20)2[21]_{FS}[21]_F[21]_S$ & ${3\over 2}^+, N(1900)?;
{5\over 2}^+,N(2000)?;$ &1950?
&$-{15\over 4}P_{00}^\pi+{1\over 4}P_{00}^{uu}$\\
&&(1908)&$-{15\over 4}P_{22}^\pi+{1\over 4}P_{22}^{uu}$\\
&&&$+{9\over 2}P_{11}^\pi+{1\over 2}P_{11}^{uu}$\\ \hline

$2(20)2[21]_{FS}[21]_F[3]_S$ & ${1\over 2}^+, N(?);
{3\over 2}^+,N(?)$&1990?
&$-{3\over 4}P_{00}^\pi-{1\over 4}P_{00}^{uu}$\\
&${5\over 2}^+,N(?);{7\over 2}^+,N(1990)?$&(1858)&
$-{3\over 4}P_{22}^\pi-{1\over 4}P_{22}^{uu}$\\
&&&$+{9\over 2}P_{11}^\pi-{1\over 2}P_{11}^{uu}$\\
\hline

$2(20)0[21]_{FS}[3]_F[21]_S$ & ${1\over 2}^+, \Delta(1910)
$ &1910&$-{3\over 4}P_{00}^\pi-{1\over 4}P_{00}^{uu}$\\
&&(1902)&$-{3\over 4}P_{20}^\pi-{1\over 4}P_{20}^{uu}$\\
&&&$+{9\over 2}P_{11}^\pi+{3\over 2}P_{11}^{uu}$\\ \hline

$2(20)2[21]_{FS}[3]_F[21]_S$ & ${3\over 2}^+, \Delta(1920);
{5\over 2}^+,\Delta(1905)$ &1912
&$-{3\over 4}P_{00}^\pi-{1\over 4}P_{00}^{uu}$\\
&&(1942)&$-{3\over 4}P_{22}^\pi-{1\over 4}P_{22}^{uu}$\\
&&&$+{9\over 2}P_{11}^\pi+{3\over 2}P_{11}^{uu}$\\
 \hline
\end{tabular}
\end{center}

\newpage
\centerline{\bf Table 7}
\vspace{0.5cm}

The structure of the $\Lambda$-hyperon
states in the $N=1,2$ bands predicted with the $SU(3)_F$
breaking
version of the chiral boson exchange interaction.
The superscripts uu and us on the $\eta$ exchange matrix
elements indicate that they apply to pair states of two
light and one light and one strange quark respectively.
The predicted energies (in MeV)
are given in the brackets under the empirical values.
For the parameter set see Table 6.
\begin{center}
\begin{tabular}{|llll|} \hline
$N(\lambda\mu)L[f]_{FS}[f]_F[f]_S$
& LS multiplet & average &$\delta M_\chi$\\
&&energy&\\ \hline
$1(10)1[21]_{FS}[111]_F[21]_S$ & ${1\over 2}^-, \Lambda(1405);
$ &
1462&$-{9\over 2}P_{00}^\pi+{1\over 2}P_{00}^{uu}-2P_{00}^{us}
-6 P_{00}^K$\\
&${3\over 2}^-,\Lambda(1520)$&(1498)
&$+{3\over 2}P_{11}^\pi-{1\over 6}P_{11}^{uu}
+{2\over 3}P_{11}^{us}+2 P_{11}^K$\\ \hline

$2(20)0[3]_{FS}[21]_F[21]_S$ & ${1\over 2}^+, \Lambda(1600)$ &
1600&$-{9\over 2} P_{00}^\pi+{1\over 2}P_{00}^{uu}-3P_{00}^K$\\
&&(1606)&$-{9\over 2}P_{20}^\pi+{1\over 2}P_{20}^{uu}-3P_{20}^K$\\
\hline

$1(10)1[21]_{FS}[21]_F[21]_S$ & ${1\over 2}^-, \Lambda(1670);
$ &
1680&$-{9\over 2}P_{00}^\pi+{1\over 2}P_{00}^{uu}
-3 P_{00}^K$\\
&${3\over 2}^-, \Lambda(1690)$
&(1629)&$+{3\over 2}P_{11}^\pi-{1\over 6}P_{11}^{uu}
-{4\over 3}P_{11}^{us}+5 P_{11}^K$\\ \hline

$1(10)1[21]_{FS}[21]_F[3]_S$ & ${1\over 2}^-, \Lambda(1800);
{3\over 2}^-,\Lambda(?);$ &
1815&$+P_{00}^{us}-3P_{00}^K$\\
&${5\over 2}^-,\Lambda(1830)$&(1756)&
$+3P_{11}^\pi-{1\over 3}P_{11}^{uu}+{1\over 3}P_{11}^{us}
+P_{11}^K$\\ \hline

$2(20)0[21]_{FS}[111]_F[21]_S$ & ${1\over 2}^+, \Lambda(1810)
$&1810&$-{9\over 4}P_{00}^\pi+{1\over 4}P_{00}^{uu}-P_{00}^{us}
-3P_{00}^K$\\
&&(1797)&$-{9\over 4}P_{20}^\pi+{1\over 4}P_{20}^{uu}
-P_{20}^{us}-3P_{20}^K$\\
&&&$+{3\over 2}P_{11}^\pi-{1\over 6}P_{11}^{uu}
+{2\over 3}P_{11}^{us}+2P_{11}^K$\\ \hline

$2(20)2[3]_{FS}[21]_F[21]_S$ & ${3\over 2}^+, \Lambda(1890);
$ &
1855&$-{9\over 2} P_{00}^\pi+{1\over 2}P_{00}^{uu}-3P_{00}^K$\\
&${5\over 2}^+,\Lambda(1820)$
&(1855)&$-{9\over 2}P_{22}^\pi+{1\over 2}P_{22}^{uu}-3P_{22}^K$\\
\hline

$2(20)0[21]_{FS}[21]_F[21]_S$&${1\over 2}^+,\Lambda(?)$&
?&$-{9\over 4}P_{00}^\pi+{1\over 4}P_{00}^{uu}-{3\over 2}P_{00}^K
$\\
&&(1872)&$-{9\over 4}P_{20}^\pi+{1\over 4}P_{20}^{uu}
-{3\over 2}P_{20}^K$\\
&&&$+{3\over 2}P_{11}^\pi-{1\over 6}P_{11}^{uu}
-{4\over 3}P_{11}^{us}+5P_{11}^K$\\ \hline

$2(20)0[21]_{FS}[21]_F[3]_S$ & ${3\over 2}^+, \Lambda(?)$&
?&$+{1\over 2}P_{00}^{us}-{3\over 2}P_{00}^K$\\
&&(1937)&$+{1\over 2}P_{20}^{us}-{3\over 2}P_{20}^K$\\
&&&$+3P_{11}^\pi-{1\over 3}P_{11}^{uu}
+{1\over 3}P_{11}^{us}+P_{11}^K$\\ \hline

$2(20)2[21]_{FS}[21]_F[3]_S$ & ${1\over 2}^+, \Lambda(?);
{3\over 2}^+,\Lambda (?);$&2020?&
$+{1\over 2}P_{00}^{us}-{3\over 2}P_{00}^K$\\
&${5\over 2}^+\Lambda(?);{7\over 2}^+,\Lambda(2020)?$&(1970)&
$+{1\over 2}P_{22}^{us}-{3\over 2}P_{22}^K$\\
&&&$+3P_{11}^\pi-{1\over 3}P_{11}^{uu}+{1\over 3}P_{11}^{us}
+P_{11}^K$\\ \hline

$2(20)2[21]_{FS}[111]_F[21]_S$ & ${3\over 2}^+, \Lambda(?);
{5\over 2}^+,\Lambda(2110)?$&
2110?&$-{9\over 4}P_{00}^\pi+{1\over 4}P_{00}^{uu}-P_{00}^{us}
-3P_{00}^K$\\
&&(2005)&$-{9\over 4}P_{22}^\pi+{1\over 4}P_{22}^{uu}
-P_{22}^{us}
-3P_{22}^K$\\
&&&$+{3\over 2}P_{11}^\pi-{1\over 6}P_{11}^{uu}
+{2\over 3}P_{11}^{us}+2P_{11}^K$\\ \hline

$2(20)2[21]_{FS}[21]_F[21]_S$ & ${3\over 2}^+,\Lambda(?);
{5\over 2}^+,\Lambda(2110)?$ &2110?
&$-{9\over 4}P_{00}^\pi+{1\over 4}P_{00}^{uu}-{3\over 2}P_{00}^K
$\\
&&(1996)&$-{9\over 4}P_{22}^\pi+{1\over 4}P_{22}^{uu}
-{3\over 2}P_{22}^K$\\
&&&$+{3\over 2}P_{11}^\pi-{1\over 6}P_{11}^{uu}
-{4\over 3}P_{11}^{us}+5P_{11}^K$\\ \hline

\end{tabular}
\end{center}

\newpage
\centerline{\bf Table 8}
\vspace{0.5cm}

The structure of the $\Sigma$-hyperon
states in the $N=1,2$ bands predicted with the $SU(3)_F$
breaking
version of the chiral boson exchange interaction.
The superscripts uu and us on the $\eta$ exchange matrix
elements indicate that they apply to pair states of two
light and one light and one strange quark respectively.
The predicted energies (in MeV)
are given in the brackets under the empirical values.
For the parameter set see Table 6.

\begin{center}
\begin{tabular}{|llll|} \hline
$N(\lambda\mu)L[f]_{FS}[f]_F[f]_S$
& LS multiplet & average &$\delta M_\chi$\\
&&energy&\\ \hline
$2(20)0[3]_{FS}[21]_F[21]_S$ & ${1\over 2}^+,\Sigma(1660)$ &
1660&$-{1\over 2} P_{00}^\pi -{1\over 6}P_{00}^{uu}
-{4\over 3}P_{00}^{us}-5P_{00}^K$\\
&&(1660)&$-{1\over 2} P_{20}^\pi -{1\over 6}P_{20}^{uu}
-{4\over 3}P_{20}^{us}-5P_{20}^K$\\
\hline

$1(10)1[21]_{FS}[21]_F[21]_S$ & ${1\over 2}^-, \Sigma(1620);
{3\over 2}^-, \Sigma(1580)$ &
1600&$-{1\over 2}P_{00}^\pi-{1\over 6}P_{00}^{uu}
-{4\over 3}P_{00}^{us}
-5 P_{00}^K$\\
&&(1667)&$+{3\over 2}P_{11}^\pi+{1\over 2}P_{11}^{uu}
+3 P_{11}^K$\\ \hline

$2(20)0[3]_{FS}[3]_F[3]_S$ & ${3\over 2}^+, \Sigma(?)$ &
?&$-{1\over 2} P_{00}^\pi-{1\over 6}P_{00}^{uu}
+{2\over 3}P_{00}^{us}- 2P_{00}^K$\\
&&(1748)&$- {1\over 2}P_{20}^\pi -{1\over 6}P_{20}^{uu}
+{2\over 3}P_{20}^{us}- 2P_{20}^K$
\\ \hline

$1(10)1[21]_{FS}[3]_F[21]_S$ & ${1\over 2}^-, \Sigma(1750)?;
{3\over 2}^-,\Sigma(?)$ &
1750?&$-{1\over 2}P_{00}^\pi-{1\over 6}P_{00}^{uu}
+{2\over 3}P_{00}^{us}
-2 P_{00}^K$\\
&&(1798)&$+{3\over 2}P_{11}^\pi+{1\over 2}P_{11}^{uu}
-2P_{11}^{us}+6 P_{11}^K$\\ \hline

$1(10)1[21]_{FS}[21]_F[3]_S$ & ${1\over 2}^-, \Sigma(1750)?;
$ &
1732&$-P_{00}^\pi-{1\over 3}P_{00}^{uu}+{1\over 3}P_{00}^{us}
-P_{00}^K$\\
&${3\over 2}^-,\Sigma(1670);{5\over 2}^-,\Sigma(1775)$&(1703)&
$+P_{11}^{us}+3P_{11}^K$\\ \hline

$2(20)2[3]_{FS}[3]_F[3]_S$&${1\over 2}^+,\Sigma(1770)?;
$&1805?&$- {1\over 2}P_{00}^\pi
-{1\over 6}P_{00}^{uu}
+{2\over 3}P_{00}^{us}-2 P_{00}^K$\\
&${3\over 2}^+,\Sigma(1840)?;{5\over 2}^+,\Sigma(?);$&(1819)&
$- {1\over 2}P_{22}^\pi -{1\over 6}P_{22}^{uu}
+{2\over 3}P_{22}^{us}- 2P_{22}^K $\\
&${7\over 2}^+,\Sigma(?)$&& \\ \hline

$2(20)2[3]_{FS}[21]_F[21]_S$&${3\over 2}^+,\Sigma(?);
{5\over 2}^+,\Sigma(1915)$&1915&
$-{1\over 2} P_{00}^\pi -{1\over 6}P_{00}^{uu}
-{4\over 3}P_{00}^{us}-5P_{00}^K$\\
&&(1897)&
$-{1\over 2} P_{22}^\pi -{1\over 6}P_{22}^{uu}
-{4\over 3}P_{22}^{us}-5P_{22}^K$\\
\hline

$2(20)0[21]_{FS}[21]_F[21]_S$ & ${1\over 2}^+, \Sigma(1880)?$
&1880?&$-{1\over 4}P_{00}^\pi-{1\over 12}P_{00}^{uu}
-{2\over 3}P_{00}^{us}
-{5\over 2}P_{00}^K$\\
&&(1906)&$-{1\over 4}P_{20}^\pi-{1\over 12}P_{20}^{uu}
-{2\over 3}P_{20}^{us}
-{5\over 2}P_{20}^K$\\
&&&$+{3\over 2}P_{11}^\pi+{1\over 2}P_{11}^{uu}+3P_{11}^K$\\ \hline

$2(20)0[21]_{FS}[21]_F[3]_S$ & ${3\over 2}^+, \Sigma(?)
$ &?&$-{1\over 2}P_{00}^\pi-{1\over 6}P_{00}^{uu}
+{1\over 6}P_{00}^{us}-{1\over 2}P_{00}^K$\\
&&(1887)&$-{1\over 2}P_{20}^\pi-{1\over 6}P_{20}^{uu}
+{1\over 6}P_{20}^{us}-{1\over 2}P_{20}^K$\\
&&&$+P_{11}^{us}+3P_{11}^K$\\ \hline

$2(20)2[21]_{FS}[21]_F[21]_S$ & ${3\over 2}^+, \Sigma (?);
{5\over 2}^+,\Sigma(?);$ &?
&$-{1\over 4}P_{00}^\pi-{1\over 12}P_{00}^{uu}-{2\over 3}P_{00}^{us}
-{5\over 2}P_{00}^K$\\
&&(2025)&$-{1\over 4}P_{22}^\pi-{1\over 12}P_{22}^{uu}
-{2\over 3}P_{22}^{us}
-{5\over 2}P_{22}^K$\\
&&&$+{3\over 2}P_{11}^\pi+{1\over 2}P_{11}^{uu}+3P_{11}^K$\\ \hline

$2(20)2[21]_{FS}[21]_F[3]_S$ & ${1\over 2}^+, \Sigma(?);
{3\over 2}^+,\Sigma(2080)?;$&2060?
&$-{1\over 2}P_{00}^\pi-{1\over 6}P_{00}^{uu}
+{1\over 6}P_{00}^{us}-{1\over 2}P_{00}^K$\\
&${5\over 2}^+,\Sigma(2070)?;
$&(1925)&$-{1\over 2}P_{22}^\pi-{1\over 6}P_{22}^{uu}
+{1\over 6}P_{22}^{us}-{1\over 2}P_{22}^K$\\
&${7\over 2}^+,\Sigma(2030)$&&$+P_{11}^{us}+3P_{11}^K$\\ \hline

$2(20)0[21]_{FS}[3]_F[21]_S$ & ${1\over 2}^+, \Sigma(?)
$ &?&$-{1\over 4}P_{00}^\pi-{1\over 12}P_{00}^{uu}
+{1\over 3}P_{00}^{us}-P_{00}^K$\\
&&(1981)&$-{1\over 4}P_{20}^\pi-{1\over 12}P_{20}^{uu}
+{1\over 3}P_{20}^{us}-P_{20}^K$\\
&&&$+{3\over 2}P_{11}^\pi+{1\over 2}P_{11}^{uu}
-2P_{11}^{us}+6P_{11}^K$\\ \hline

$2(20)2[21]_{FS}[3]_F[21]_S$ & ${3\over 2}^+, \Sigma(2080)?;
$ &2075?
&$-{1\over 4}P_{00}^\pi-{1\over 12}P_{00}^{uu}
+{1\over 3}P_{00}^{us}-P_{00}^K$\\
&${5\over 2}^+,\Sigma(2070)?$
&(2016)&$-{1\over 4}P_{22}^\pi-{1\over 12}P_{22}^{uu}
+{1\over 3}P_{22}^{us}-P_{22}^K$\\
&&&$+{3\over 2}P_{11}^\pi+{1\over 2}P_{11}^{uu}
-2P_{11}^{us}+6P_{11}^K$\\ \hline

\end{tabular}
\end{center}

\newpage
\centerline{\bf Table 9}
\vspace{0.5cm}

The structure of the $\Xi$-hyperon
states in the $N=1,2$ bands predicted with the $SU(3)_F$
breaking
version of the chiral boson exchange interaction. The
superscripts us and ss on the $\eta$-exchange matrix
elements indicate that the interaction acts in pair
states of one light and onw strange and two strange quarks
respectively.
The predicted energies (in MeV)
are given in the brackets under the empirical values.
For the parameter set see Table 6.
\begin{center}
\begin{tabular}{|llll|} \hline
$N(\lambda\mu)L[f]_{FS}[f]_F[f]_S$
& LS multiplet & average &$\delta M_\chi$\\
&&energy&\\ \hline
$2(20)0[3]_{FS}[21]_F[21]_S$ & ${1\over 2}^+,\Xi(?)$ &
?&$-{4\over 3}P_{00}^{us}-{2\over 3}P_{00}^{ss}-5P_{00}^K$\\
&&(1798)&$-{4\over 3}P_{20}^{us}-{2\over 3}P_{20}^{ss}
-5P_{20}^K$\\ \hline

$1(10)1[21]_{FS}[21]_F[21]_S$ & ${1\over 2}^-, \Xi(?);
{3\over 2}^-, \Xi(?)$ &
?&$-{4\over 3} P_{00}^{us}-{2\over 3}P_{00}^{ss}- 5P_{00}^K$\\
&&(1758)&$+2P_{11}^{ss}+3 P_{11}^K$\\ \hline

$2(20)0[3]_{FS}[3]_F[3]_S$ & ${3\over 2}^+, \Xi(?)$ &
?&${2\over 3}P_{00}^{us}-{2\over 3}P_{00}^{ss}-2 P_{00}^K$\\
&&(1886)& ${2\over 3}P_{20}^{us}-{2\over 3}P_{20}^{ss}
-2P_{20}^K$ \\ \hline

$1(10)1[21]_{FS}[3]_F[21]_S$ & ${1\over 2}^-, \Xi(?);
$ &
1820?&${2\over 3}P_{00}^{us}-{2\over 3}P_{00}^{ss}-2P_{00}^K$\\
&${3\over 2}^-,\Xi(1820)?$
&(1889)&$-2P_{11}^{us}+2P_{11}^{ss}+6P_{11}^K$\\ \hline

$1(10)1[21]_{FS}[21]_F[3]_S$ & ${1\over 2}^-, \Xi(?);
{3\over 2}^-,\Xi(?);$ &
?&$+{1\over 3} P_{00}^{us}-{4\over 3}P_{00}^{ss}-P_{00}^K$\\
&${5\over 2}^-,\Xi(?)$&(1849)&$+P_{11}^{us}+3 P_{11}^K$\\
\hline

$2(20)2[3]_{FS}[3]_F[3]_S$&${1\over 2}^+,\Xi(?);
{3\over 2}^+,\Xi(?);$&?&${2\over 3}P_{00}^{us}-{2\over 3}P_{00}^{ss}
-2P_{00}^K$\\
&${5\over 2}^+,\Xi(?);{7\over 2}^+,\Xi(?)$&(1947)&
${2\over 3}P_{22}^{us}-{2\over 3}P_{22}^{ss}-2P_{22}^K$\\ \hline

$2(20)2[3]_{FS}[21]_F[21]_S$&${3\over 2}^+,\Xi(?);
{5\over 2}^+,\Xi(?)$&?&$-{4\over 3}P_{00}^{us}
-{2\over 3}P_{00}^{ss}-5P_{00}^K$\\
&&(2025)&$-{4\over 3}P_{22}^{us}-{2\over 3}P_{22}^{ss}
-5P_{22}^K$\\ \hline

$2(20)0[21]_{FS}[21]_F[21]_S$ & ${1\over 2}^+, \Xi(?)$
&?&$-{2\over 3}P_{00}^{us}-{1\over 3}P_{00}^{ss}
-{5\over 2}P_{00}^K$\\
&&(1994)&$-{2\over 3}P_{20}^{us}-{1\over 3}P_{20}^{ss}
-{5\over 2}P_{20}^K$\\
&&&$+2 P_{11}^{ss}+3 P_{11}^K$\\ \hline

$2(20)0[21]_{FS}[21]_F[3]_S$ & ${3\over 2}^+, \Xi(?)
$ &?&${1\over 6}P_{00}^{us}-{2\over 3}P_{00}^{ss}
-{1\over 2}P_{00}^K$\\
&&(2026)&${1\over 6}P_{20}^{us}-{2\over 3}P_{20}^{ss}
-{1\over 2}P_{20}^K$\\
&&&$+P_{11}^{us}+3P_{11}^K$\\ \hline

$2(20)2[21]_{FS}[21]_F[21]_S$ & ${3\over 2}^+, \Xi (?);
{5\over 2}^+,\Xi(?);$ &?
&$-{2\over 3}P_{00}^{us}-{1\over 3}P_{00}^{ss}-{5\over 2}P_{00}^K$\\
&&(2107)&$-{2\over 3}P_{22}^{us}-{1\over 3}P_{22}^{ss}
-{5\over 2}P_{22}^K$\\
&&&$+2 P_{11}^{ss}+3 P_{11}^K$\\ \hline

$2(20)2[21]_{FS}[21]_F[3]_S$ & ${1\over 2}^+, \Xi(?);
{3\over 2}^+,\Xi(?);$&?
&${1\over 6}P_{00}^{us}-{2\over 3}P_{00}^{ss}
-{1\over 2}P_{00}^K$\\
&${5\over 2}^+,\Xi(?);{7\over 2}^+,\Xi(?)
$&(2053)&${1\over 6}P_{22}^{us}-{2\over 3}P_{22}^{ss}
-{1\over 2}P_{22}^K$\\
&&&$+P_{11}^{us}+3P_{11}^K$\\ \hline

$2(20)0[21]_{FS}[3]_F[21]_S$ & ${1\over 2}^+, \Xi(?)
$ &?&${1\over 3}P_{00}^{us}-{1\over 3}P_{00}^{ss}-P_{00}^K$\\
&&(2069)&${1\over 3}P_{20}^{us}-{1\over 3}P_{20}^{ss}-P_{20}^K$\\
&&&$-2P_{11}^{us}+2P_{11}^{ss}+6P_{11}^K$\\ \hline

$2(20)2[21]_{FS}[3]_F[21]_S$ & ${3\over 2}^+, \Xi(?);
{5\over 2}^+,\Xi(?)$ &?
&${1\over 3}P_{00}^{us}-{1\over 3}P_{00}^{ss}-P_{00}^K$\\
&&(2099)&${1\over 3}P_{22}^{us}-{1\over 3}P_{22}^{ss}-P_{22}^K$\\
&&&$-2P_{11}^{us}+2P_{11}^{ss}+6P_{11}^K$\\  \hline
\end{tabular}
\end{center}
\newpage
\centerline{\bf Table 10}
\vspace{0.5cm}

The structure of the $\Omega^-$ hyperon
states in the $N=1,2$ bands predicted with the $SU(3)_F$
breaking
version of the chiral boson exchange interaction.
The predicted energies (in MeV)
are given in the brackets under the empirical values.
For the parameter set see Table 6.
\begin{center}
\begin{tabular}{|llll|} \hline
$N(\lambda\mu)L[f]_{FS}[f]_F[f]_S$
& LS multiplet & average &$\delta M_\chi$\\
&&energy&\\ \hline

$2(20)0[3]_{FS}[3]_F[3]_S$ & ${3\over 2}^+, \Omega^-(?)$ &
?&$- 2P_{00}^{ss}- 2P_{20}^{ss} $\\
&&(2020)& \\ \hline

$1(10)1[21]_{FS}[3]_F[21]_S$ & ${1\over 2}^-, \Omega^-(?);
{3\over 2}^-,\Omega^-(?)$ &
?&$-2P_{00}^{ss}+6P_{11}^{ss}$\\
&&(1991)&\\ \hline

$2(20)2[3]_{FS}[3]_F[3]_S$&${1\over 2}^+,\Omega^-(?);
{3\over 2}^+,\Omega^-(?);$&?&$-2P_{00}^{ss}-2P_{22}^{ss}$\\
&${5\over 2}^+,\Omega^-(?);{7\over 2}^+,\Omega^-(?)$&(2068)&
\\ \hline

$2(20)0[21]_{FS}[3]_F[21]_S$ & ${1\over 2}^+, \Omega^-(?)
$ &?&$-P_{00}^{ss}-P_{20}^{ss}+6 P_{11}^{ss}$\\
&&(2166)&\\ \hline

$2(20)2[21]_{FS}[3]_F[21]_S$ & ${3\over 2}^+, \Omega^-(?);
{5\over 2}^+,\Omega^-(?)$ &?
&$-P_{00}^{ss}-P_{22}^{ss}+6P_{11}^{ss}$\\
&&(2190)&\\ \hline
\end{tabular}
\end{center}

\newpage
\centerline{\bf Table 11}
\vspace{1cm}

The contributions to the baryon energies from the interactions
(3.3),(8.1) and(9.1), with inclusion of the nondiagonal matrix elements
of the spin-orbit and tensor interactions. The terms
$\delta M_\chi$ are the contributions to the corresponding
states from the spin-spin interaction(3.3), which are
listed in Tables 6-9. The net spin-orbit ($V_{11}^*$) and tensor
interaction ($T_{11}^*$) matrix elements for the different
sectors of the baryon spectrum are listed in Table 12. The
$[3]_F$ states are absent here as both tensor and spin-orbit
forces do not contribute in this case.

\vspace{1cm}

\begin{center}
\begin{tabular}{|lll|} \hline
&&\\
$[f]_{FS}[f]_F[f]_S$ & \,\,\,\,\,Potential & matrix  \\
&&\\ \hline
&&\\
$[21]_{FS}[21]_F[21]_S:\frac{1}{2}^-$
& $\delta M_\chi([21]_F[21]_S)-V_{11}^*$ & $-8T_{11}^*+
\frac{1}{2}V_{11}^*$ \\
 &  &  \\
$[21]_{FS}[21]_F[3]_S:\frac{1}{2}^-$
 & $-8T_{11}^*+\frac{1}{2}V_{11}^*$
& $\delta M_\chi([21]_F[3]_S)+8T_{11}^*-\frac{5}{2}V_{11}^*$\\
&&\\ \hline
&&\\
$[21]_{FS}[21]_F[21]_S:\frac{3}{2}^-$
& $\delta M_\chi([21]_F[21]_S)+\frac{1}{2}V_{11}^*$ &
$\frac{4\sqrt{10}}{5}T_{11}^*+\frac{\sqrt{10}}{4}V_{11}^*$\\
 & & \\
$[21]_{FS}[21]_F[3]_S:\frac{3}{2}^-$ &$\frac{4\sqrt{10}}{5}T_{11}^*+
\frac{\sqrt{10}}{4}V_{11}^*$ & $\delta M_\chi([21]_F[3]_S)
-\frac{32}{5}T_{11}^*-
V_{11}^*$\\
&&\\ \hline
&&\\
$[21]_{FS}[111]_F[21]_S:\frac{1}{2}^-$
 & $\delta M_\chi([111]_F[21]_S)-2V_{11}^*$ & \\
 & & \\ \hline
&&\\
$[21]_{FS}[111]_F[21]_S:\frac{3}{2}^-$ & $\delta M_\chi([111]_F[21]_S)
+V_{11}^*$ & \\
 & & \\ \hline
&&\\
$[21]_{FS}[21]_F[3]_S:
\frac{5}{2}^-$ & $\delta M_\chi([21]_F[31]_S)+\frac{8}{5}T_{11}^*+
\frac{3}{2}V_{11}^*$ & \\
 & & \\ \hline
\end{tabular}
\end{center}

\newpage
\centerline{\bf Table 12}
\vspace{0.5cm}

The contributions to the net spin-orbit ($V_{11}^*$) and
tensor ($T_{11}^*$) interaction matrix elements from the
different exchange interactions in eqs. (8.1) and
(9.1) for the $N=L=1$ baryon states. The
the superscripts uu and us on the matrix elements of the
$\omega$-exchange-like
spin-orbit interaction and of the $\eta$-exchange tensor
interaction indicate that they apply
to pair states of two light and one light and one strange quark
respectively.
\begin{center}
\begin{tabular}{|l|l|l|} \hline
&&\\
& $V_{11}^*$ &$T_{11}^*$\\
&&\\ \hline
&&\\
$N$ & $-V_{11}^S+3V_{11}^\rho-{1\over 3}V_{11}^{uu}$&
$-3T_{11}^{\pi}+{1\over 3}T_{11}^{uu}$\\
&&\\ \hline
&&\\
$\Lambda([111]_F)$&$-V_{11}^S+V_{11}^\rho+{4\over 3}V_{11}^{K^*}
-{1\over 9}V_{11}^{uu}+{4\over 9}V_{11}^{us} $&
$-T_{11}^{\pi}-{4\over 3}T_{11}^{K}+{1\over 9}T_{11}^{uu}-
{4\over 9}T_{11}^{us}$\\
&&\\ \hline
&&\\
$\Lambda([21]_F)$&$-V_{11}^S+2V_{11}^\rho+{2\over 3}V_{11}^{K^*}
-{2\over 9}V_{11}^{uu}+{2\over 9}V_{11}^{us} $&
$-2T_{11}^{\pi}-{2\over 3}T_{11}^{K}+{2\over 9}T_{11}^{uu}-
{2\over 9}T_{11}^{us}$\\
&&\\ \hline
&&\\
$\Sigma$&$-V_{11}^S+2V_{11}^{K^*}+{2\over 3}V_{11}^{us}$&
$-2T_{11}^{K}-{2\over 3}T_{11}^{us}$\\
&&\\ \hline
&&\\
$\Xi$&$-V_{11}^S+2V_{11}^{K^*}+{2\over 3}V_{11}^{us}$&
$-2T_{11}^{K}-{2\over 3}T_{11}^{us}$\\
&&\\ \hline
\end{tabular}
\end{center}

\newpage

\centerline{\bf Table 13}
\vspace{0.5cm}

Magnetic moments of the baryon octet (in nuclear magnetons). Column
IA contains the quark model impulse approximation expressions,
column "exp" the experimental values, column I the impulse
approximation predictions, column II the exchange current contribution
with
$<\varphi_{000}(\vec r_{12})
|\tilde{V}_\pi(r_{12})|\varphi_{000}(\vec r_{12})> =  - 0.018$
and
$<\varphi_{000}(\vec r_{12})
|\tilde{V}_K(r_{12})|\varphi_{000}(\vec r_{12})> =  0.03$
and column III the net predictions. All magnetic moments are given
in nuclear magnetons.
\vspace{1.5cm}

\begin{center}
\begin{tabular}{|l|l|l|l|l|l|} \hline
 & IA & exp & I & II & III\\ \hline
&&&&& \\
$p$ & ${m_N\over m_u}$ & +2.79 & +2.76 & +0.07 & +2.83\\
&&&&& \\
$n$ & $-{2\over 3}{m_N\over m_u}$ & --1.91 & --1.84 & --0.07 & --1.91\\
&&&&& \\
$\Lambda$ & $-{1\over 3}{m_N\over m_s}$ & --0.61 & --0.67 & +0.06 &
--0.61\\
&&&&& \\
$\Sigma^+$ & ${8\over 9}{m_N\over m_u}+{1\over 9}{m_N\over m_s}$ &
+2.42 & +2.68 & --0.12 & +2.56\\
&&&&& \\
$\Sigma^0$ & ${2\over 9} {m_N\over m_u}+{1\over 18}{m_N\over m_s}$ & ?
& +0.72 & --0.06 & +0.66\\
&&&&& \\
$\Sigma^0\rightarrow \Lambda$ & $-{1\over \sqrt{3}}{m_N\over m_u}$ &
$\vert$1.61$\vert$ & --1.59 & +0.01 & --1.58\\
&&&&& \\
$\Sigma^-$ & $-{4\over 9}{m_N\over m_u}+{1\over 9}{m_N\over m_s}$ &
--1.16 & --1.00 & 0 & --1.00\\
&&&&& \\
$\Xi^0$ & $-{2\over 9}{m_N\over m_u}-{4\over 9}{m_N\over m_s}$ &
--1.25 & --1.51 & +0.12 & --1.39\\
&&&&& \\
$\Xi^-$ & ${1\over 9}{m_N\over m_u}-{4\over 9}{m_N\over m_s}$ & --0.65
& --0.59 & 0 & --0.59\\
&&&&& \\ \hline
\end{tabular}
\end{center}
\end{document}